\documentclass[conf]{new-aiaa} %for conference papers
\usepackage[utf8]{inputenc}
\usepackage{textcomp}
\usepackage{multicol}
\usepackage{lscape}

\usepackage{cleveref}
\usepackage{algorithm}
\usepackage{algpseudocode}
\usepackage{graphicx}
\usepackage{amsmath}
\usepackage[version=4]{mhchem}
\usepackage{siunitx}
\usepackage{longtable,tabularx}
\usepackage{xcolor} % font colors
\usepackage{wrapfig} % wrap figures
\usepackage{subcaption} % for subfigures
\usepackage{array}
\newcolumntype{L}[1]{>{\raggedright\let\newline\\\arraybackslash\hspace{0pt}}m{#1}}
\newcolumntype{C}[1]{>{\centering\let\newline\\\arraybackslash\hspace{0pt}}m{#1}}
\newcolumntype{R}[1]{>{\raggedleft\let\newline\\\arraybackslash\hspace{0pt}}m{#1}}

\usepackage{tikz}
\usetikzlibrary{shapes}

\usepackage{nomencl}
\makenomenclature

\title{Systems Theoretic Process Analysis of a \\Run Time Assured Neural Network Control System\footnote{Approved for Public Release. Case Number AFRL-2022-2453.}}

\author{Kerianne L. Hobbs \footnote{Safe Autonomy Lead, Autonomy Capability Team 3, 2241 Avionics Circle, AIAA Senior Member.} and 
Benjamin K. Heiner\footnote{Reinforcement Learning Lead Engineer, Autonomy Capability Team 3, 2241 Avionics Circle.}}
\affil{Air Force Research Laboratory, Wright-Patterson AFB, OH, 45433}
\author{Lillian Busse\footnote{Student, Aerospace Engineering, 2600 Clifton Ave, AIAA Member.}}
\affil{University of Cincinnati, Cincinnati, OH, 45221}
\author{Kyle Dunlap\footnote{AI Scientist, 4035 Colonel Glenn Hwy, AIAA Member.}}
\affil{Parallax Advanced Research, Beavercreek, OH 45431}
\author{Jonathan Rowanhill\footnote{President, 415 Cranberry Lane.} and Ashlie B. Hocking\footnote{Principal Scientist, 1173 Colbert Street.}}
\affil{Dependable Computing, Charlottesville, VA, 22901}
\author{Aditya Zutshi\footnote{Research Engineer, 421 SW 6th Ave.} }
\affil{Galois, Portland, OR, 97204}

\begin{document}

\maketitle

\begin{abstract}
This research considers the problem of identifying safety constraints and developing Run Time Assurance (RTA) for Deep Reinforcement Learning (RL) Tactical Autopilots that use neural network control systems (NNCS). This research studies a specific use case of an NNCS performing autonomous formation flight while an RTA system provides collision avoidance and geofence assurances. First, Systems Theoretic Accident Models and Processes (STAMP) is applied to identify accidents, hazards, and safety constraints as well as define a functional control system block diagram of the ground station, manned flight lead, and surrogate unmanned wingman. Then, Systems Theoretic Process Analysis (STPA) is applied to the interactions of the the ground station, manned flight lead, surrogate unmanned wingman, and internal elements of the wingman aircraft to identify unsafe control actions, scenarios leading to each, and safety requirements to mitigate risks. This research is the first application of STAMP and STPA to an NNCS bounded by RTA. 
\end{abstract}

%\tableofcontents
\section*{Nomenclature}

%\noindent(Nomenclature entries should have the units identified)

{\renewcommand\arraystretch{1.0}
\noindent\begin{longtable*}{@{}l @{\quad=\quad} l@{}}
AA & Aspect Angle \\
AFSC & Altitude Follow Stack Criteria \\
AISL & Altitude Independent Slant Range \\
CONOPS & Concept of Operations\\
CCS & Common Cause Scenario \\
ENU & East-North-Up (reference frame)\\
EPM & Envelope Protection Monitor\\
FL & Follow Length \\
HMI & Human-Machine Interface \\
ISR & Intelligence, Surveillance, and Reconnaissance\\
JEC & Jet Wash Exclusion Zone Criteria\\
LL & Lead Length\\
LVC & Live, Virtual, Constructive \\
MUMT & Manned-Unmanned Teaming\\
NNCS & Neural Network Control System\\
PPO & Proximal Policy Optimization\\
RL & Reinforcement learning \\
RTA & Run Time Assurance\\
SFC & Smooth Flight Criteria\\
SSC & Soft Safety Constraints \\
STAMP & Systems Theoretic Accident Models and Processes \\
STPA & Systems Theoretic Process Analysis \\
TMP & Test Management Program \\
TNS & Tail Nose Separation \\
TNSC & Tail Nose Separation Criteria\\
TPS & Test Pilot School\\
UCA & Unsafe Control Action \\
VFR & Visual Flight Rules \\
VISTA &Variable Stability In-flight Simulator Test Aircraft \\
VMC & Visual Meteorological Conditions \\
VSS & Variable Stability System or VISTA Simulation Systems\\
%$A$ & amplitude of oscillation \\
%$a$ & cylinder diameter \\
%$C_p$& pressure coefficient \\
%$Cx$ & force coefficient in the \textit{x} direction \\
%$Cy$ & force coefficient in the \textit{y} direction \\
%c & chord \\
%d$t$ & time step \\
%$Fx$ & $X$ component of the resultant pressure force acting on the vehicle \\
%$Fy$ & $Y$ component of the resultant pressure force acting on the vehicle \\
%$f, g$ & generic functions \\
%$h$ & height \\
%$i$ & time index during navigation \\
%$j$ & waypoint index \\
%$K$ & trailing-edge (TE) nondimensional angular deflection rate\\
%$\Theta$ & boundary-layer momentum thickness\\
%$\varrho$ & density\\
%\multicolumn{2}{@{}l}{Subscripts}\\
%cg & center of gravity\\
%$G$ & generator body\\
%iso & waypoint index
\end{longtable*}}

\section{Introduction}
\lettrine{W}{ithin} the last few years, reinforcement learning (RL) has begun to demonstrate better than human performance in high dimensional state spaces such as the game of Go~\cite{silver2016mastering,silver2017mastering}, real time strategy games such as StarCraft II~\cite{vinyals2019grandmaster}, and in military engagement scenarios such as the AlphaDogfight~\cite{pope2021hierarchical}. In aviation, RL has the potential to improve air-based wildfire monitoring \cite{julian2019distributed}, urban air mobility traffic deconfliction \cite{deniz2022multi}, air traffic flow \cite{crespo2012reinforcement}, landings \cite{tang2020deep},  maintenance scheduling \cite{andrade2021aircraft}, 
acrobatic maneuvering \cite{clarke2020deep}, control system designs \cite{konatala2021reinforcement,waslander2005multi,bohn2019deep,tandale2004preliminary,de2019reinforcement,woodbury2014autonomous}, terminal area operations and planning \cite{balakrishna2010accuracy}, and other areas. However, one of the key challenges in developing and using neural network control systems (NNCS) on full scale aircraft is providing sufficient verification evidence that the operation will be safe. Applying traditional verification to autonomous control system may require hundreds of billions of hours of testing to assure safety \cite{kalra2016driving}, which is intractably time consuming and expensive.

Run Time Assurance (RTA) \cite{schierman2020runtime,hobbs2021run} is an approach to assuring safety of complex control systems like NNCS, by monitoring the system state at run time and intervening when necessary to assure safety. This project applies the Systems Theoretic Accident Models and Processes (STAMP) and Systems Theoretic Process Analysis (STPA) to a hypothetical flight test given the moniker of ``Have Leash'' to identify design requirements prior to system development. Have Leash is envisioned to test an NNCS primary controller bounded by an RTA system. The NNCS will be designed to command an autonomous wingman aircraft to rejoin on a manned flight lead \cite{AETCMAN11248}, where the two aircraft begin separated and end in a formation flight. The RTA system monitors the wingman aircraft state, lead aircraft state, and output of the NNCS and intervenes when necessary to provide collision avoidance and geofence functions. Specialized test aircraft are considered which feature a human safety pilot on board that can take control at any time, as well as a set of safety trips that define a constrained portion of the test aircraft's flight envelope.

The contributions of this work are as follows:
\begin{itemize}
    \item First application of STAMP and STPA to an NNCS, bounded by RTA.
    \item First functional control system block diagram design of a combination of an Envelope Protection Monitor (EPM) and RTA (collision avoidance, geofence) for flight testing NNCS on full scale aircraft.
    \item List of accidents, hazards, safety constraints, soft safety constraints (SSCs), common cause scenarios (CCSs), unsafe control actions (UCAs), scenarios, and safety requirements to inform verification and validation efforts to assure safety of an NNCS bounded by RTA.
\end{itemize}

The content of this paper is as follows. In Section \ref{Related Work}, this research is placed in context with other related work. In Section \ref{Analysis Process}, an overview of STAMP and STPA analysis techniques, RL, and RTA is provided. In Section \ref{Problem Statement}, the problem statement for the example use case is presented. In Section \ref{Results}, STAMP and STPA are conducted and results are presented. Finally in Section \ref{Conclusion}, conclusions and recommendations are provided.

%%%%%%%%%%%%%%%%%%%%% Related Work %%%%%%%%%%%%%%%%%%%%%%%%%%%%%%%%%%%%%%%%%%%%%
\section{Related Work} \label{Related Work}
Several previous research programs applied STAMP and STPA to Manned-Unmanned Teaming (MUMT) of aircraft, where a manned flight lead is teamed with unmanned wingman aircraft, or hazard analysis of urban air mobility \cite{graydon2020guidance}. However, none specifically considered RTA intervening to assure safety of an NNCS. This section discusses similar research in the Have Raider project \cite{montes2016using} as well as general MUMT missions \cite{hobbs2018early,robertson2019systems}. The accidents, hazards, safety constraints, functional control structure block diagrams, and UCAs from these three previous works were taken into consideration in this work. %Further details are offered in later sections.

In 2015, STAMP and STPA were applied to the Have Raider Test Pilot School (TPS) Test Management Program (TMP) flight test \cite{montes2016using}, which have the following similarities and differences to the flight test proposed in this manuscript:

%. Similar to the autonomous rejoin task in this manuscript, the objective of the Have Raider TPS TMP was to demonstrate the ability of the Autonomous Flight Controller (AFC) in an autonomous wingman to rejoin with a manned flight lead, fly in formation with the lead, loiter, and fly a route. 
\begin{itemize}
\item Both programs are  designed to test an autonomous rejoin controller. Have Raider's test system utilized traditional control theory while this manuscript considers an NNCS trained with RL.
\item Both tests use an aircraft with a Variable Stability System (VSS). Have Raider used the Variable Stability In-flight Simulator Test Aircraft (VISTA).
\item Both systems were equipped with a midair collision avoidance system. Have Raider used the Automatic Air Collision Avoidance System (Auto ACAS); however it only recorded when Auto ACAS would have activated a recovery maneuver and the primary collision avoidance provider was the human safety pilot. By contrast, this manuscript investigates collision avoidance and geofence based on control barrier functions \cite{ames2016control,hobbs2021run}). 
\end{itemize}

A generic MUMT scenario was analyzed in \cite{hobbs2018early}, and later expanded in \cite{robertson2019systems}. Both works developed and presented a MUMT Concept of Operations (CONOPS) with eight mission phases that were then analyzed with STAMP and STPA to identify safety requirements. While the mission was not specified in \cite{hobbs2018early}, a MUMT intelligence, surveillance, and reconnaissance (ISR) mission or an air to air combat scenario were considered in \cite{robertson2019systems}.  % and specified in the Architecture Analysis \& Design Language (AADL). The CONOPS included a theoretical MUMT mission with one manned aircraft and two unmanned support ``wingman'' aircraft in 8 mission phases. Each phase had a separate STAMP diagram that included some combination of mission planning, the manned flight lead aircraft, the team lead (human on the manned aircraft), the unmanned wingman aircraft, the ground station, the crew chief, and Air Traffic Control. Another unique element of this research was the inclusion of rationale for each accident, hazard, and safety constraint. 
%
%Building on \cite{hobbs2018early}, a follow-on Master's Thesis project \cite{robertson2019systems} thesis sought to demonstrate the effectiveness of STPA as an approach to autonomous systems hazard assessment. STPA was applied to generate safety and security requirements a generic MUMT intelligence, surveillance, and reconnaissance (ISR) mission or an air to air combat scenario. Rather than creating mission phase-specific STAMP diagrams, one unified STAMP diagram was considered. A unique contribution of this project was the use of casual scenario generation, which takes STPA a step further than previous works to theorize the root cause of UCAs. The causal scenarios can be used to generate specific safety requirements. 
While the rejoin task in this manuscript differs from the tasks in previous work, this manuscript will similarly use a single STAMP diagram to describe the system, a CONOPS to inform the safety analysis, and causal scenarios to inform safety requirements.

%%%%%%%%%%%%%%%%%%%%% Preliminaries %%%%%%%%%%%%%%%%%%%%%%%%%%%%%%%%%%%%%%%%%%%%%
\section{Analysis Process} \label{Analysis Process}
This section provides background information on the STAMP and STPA techniques applied to identify safety requirements in this research, as well as an introduction to RL and RTA.

\subsection{Systems Theoretic Accident Models and Processes}

STAMP is a model of accident causation technique based on systems theory and the idea that all of the parts of a system are interconnected and their behavior affects one another \cite{leveson2011engineering}. %STAMP and its processes revolve around the idea that accidents can be prevented by controlling system behavior, not just preventing failure. STAMP treats safety like a dynamic control problem, meaning control is considered in the context of feedback loops with control actions and feedback, as shown in Figure. 
The basic elements of STAMP are accidents, hazards, and safety constraints as well as a hierarchical safety control structure block diagram. 
Accidents, hazards, and safety constraints are often listed in priority order and are sourced from applicable government regulations, eliciting stakeholder expectations, and studying previous work. Accidents in STAMP are any unacceptable loss to the stakeholders, which can include human life or injury, property damage, environmental pollution, mission loss, and other losses. Hazards in STAMP are defined with respect to relationships internal to the system or relationships with the system and the environment. Hazards under the right environmental conditions lead to a loss. STAMP uses fewer than a dozen high-level hazards \cite{leveson2011engineering}. Safety constraints are generated based on a hazard and are used as high-level safety requirements that shall be met by the system.

\begin{figure}[hbt!]
\centering
\includegraphics[width =.35\textwidth,trim={5.5cm 17.5cm 10cm 4.5cm},clip]{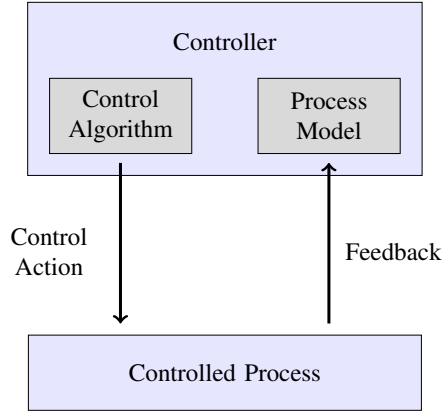} 
\caption{Components of a STAMP functional control block diagram.}
\label{fig:STAMPGeneric}
\end{figure}

%Control structure block diagrams are developed by identifying the responsibilities of multiple controllers and the controlled processes throughout the system, as well as inputs and outputs of each. 
As pictured in Figure \ref{fig:STAMPGeneric}, the elements of a control structure are controllers, controlled processes, control actions, and feedback. Controllers and controlled processes are separate functional elements. Feedback is information about the state of the controlled process, such as position or sensor data. A control action is any command that is given to a controlled process. In some models, controller subsystems in the form of control algorithms and process models may be considered. In traditional controls the process model might be some form of estimation while the control algorithm is any form of traditional controller designs. When the controller is a human, the process model might be the human's mental model of the controlled process and the control algorithm might be how the human decides to interact with the controlled process. 
Hierarchical control structures place components with higher authority higher above the processes they control. A controlled process at one level can be the controller of a process with lower authority.

\subsection{Systems Theoretic Process Analysis}\label{sec:STPATechnicalApproach}

\begin{figure}[hbt!]
\centering
\includegraphics[width =0.8\textwidth]{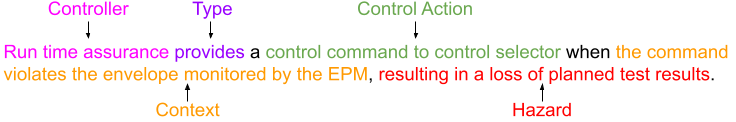}
\caption{Components of an unsafe control action.}
\label{fig:UCA_Example}
\end{figure}

Systems Theoretic Process Analysis (STPA) expands hazard analysis beyond electromechanical components failures to include design errors, component interaction accidents, human decision-making errors, and social, organizational, and management factors that may contribute to accidents \cite{leveson2011engineering}. STPA analyzes the functional control structure to find UCAs and develops causal scenarios as part of a process to identify requirements to increase system safety. Causal scenarios show how UCAs can happen and are used to develop specific safety recommendations that could be implemented in the design. The optional last step is to determine what requirements can be realistically implemented in the design and to recommend them to eliminate UCAs. UCAs are control actions or feedback that are unsafe in one of the following ways:
\begin{itemize}
\item Provided (in an inappropriate context)
\item Not provided (in a context where it should be)
\item Duration (a continuous control action is provided for too long or too short a duration)
\item Timing (a control action is provided too early or too late)
\end{itemize}
If a control action is unsafe when paired with one of the types, it becomes part of an UCA and a rationale is written to explain why it is unsafe. Written UCAs follow the form shown in Fig. \ref{fig:UCA_Example}.
%
%After developing UCAs, the next step of STPA is causal scenario and requirement generation. Generating causal scenarios can be done alongside the UCA generation or it can be a separate process done after UCA generation. Causal scenarios and requirements can be identified by considering four main types of scenarios: The command is not followed or followed inadequately, the controller makes an inappropriate decision, the controller gets inadequate feedback or input, or there is inadequate process behavior.  \cite{robertson2019systems}.

\subsection{Reinforcement Learning}

RL is a form of machine learning (ML) based on the concept of operant conditioning where desirable behavior is reinforced with rewards (positive reinforcement) and bad behavior is penalized (negative reinforcement), and the agent learns which actions to take to maximize reward through trial and error \cite{sutton2018reinforcement}. Over the course of thousands to millions of simulated interactions, the agent refines it's a policy $\pi$, which is a mapping of the environment states $s$ or observations $o$ to actions $a$. This paper assumes the use of proximal policy optimization (PPO) gradient RL \cite{schulman2017proximal}, which is a policy gradient technique that uses two neural networks: an \textit{actor} and a \textit{critic}. The actor estimates the policy while the critic estimates expected reward \cite{konda2000actor}. Additional examples of applying PPO RL to aerospace control may be found in \cite{ravaioli2022safe}. %The RL model is defined via agent-environment interactions which is represented in Figure \ref{fig:RLDiagram}.
%
%\begin{itemize}
%Environment
%\end{itemize}
%
\begin{figure}[hbt!]
\centering
\includegraphics[width=0.25\columnwidth]{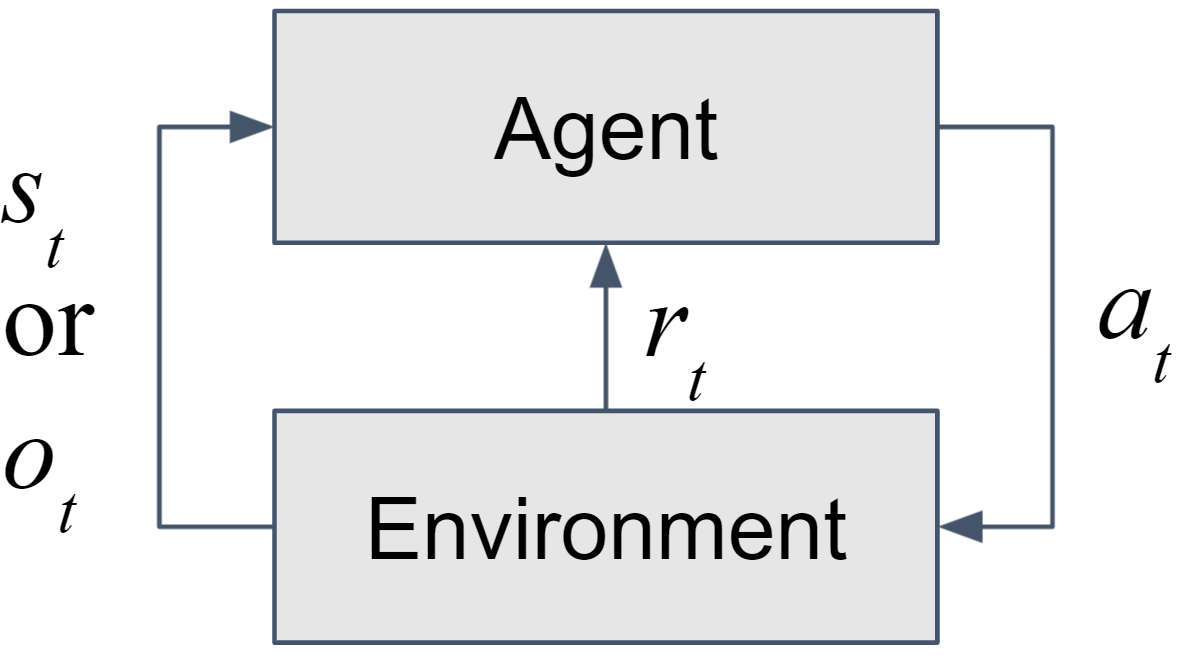}
\caption{Reinforcement learning feedback loop.}
\label{fig:RLDiagram}
\end{figure}
%
%RL is used in this research to train NNCS to perform aerospace control tasks. 
The RL training process is depicted in Figure \ref{fig:RLDiagram}. At fixed time steps, an agent (the NNCS) samples the state $s_t$ or and observation $o_t$ of its environment by sending control actions $a_t$ provided by a set of policies to the environment (the aircraft under control). The environment's state at the next timestep $s_{t+1}$, or an observation of the state by sensors $o_{t+1}$, and reward $r_{t+1}$ is then provided back to the agent. The RL algorithm inside the agent to uses the input [$s_t$,$r_t$] to update the weights and biases in the neural network to develop and optimal policy $\pi*$ that maximizes the reward.  

\subsection{Run Time Assurance}
\begin{figure}[hbt!]
\centering
\includegraphics[width=0.4\columnwidth]{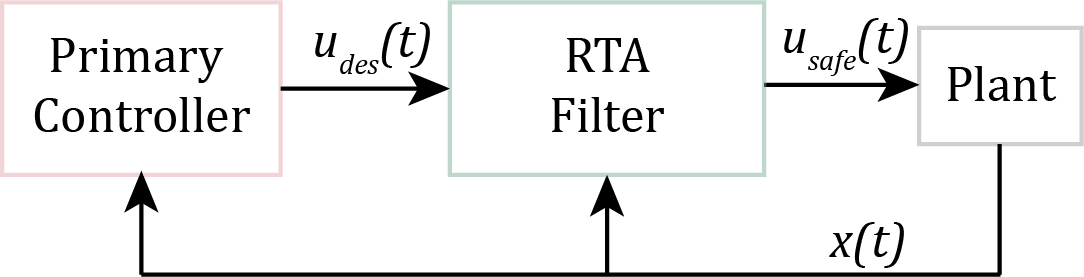}
\caption{Run time assurance monitoring a primary controller, such as a neural network control system.}
\label{fig:RTARLDiagram}
\end{figure}

RTA can be thought of as a safety filter on the autonomous control output \cite{hobbs2021run}. RTA monitors the state of the plant (and desired output of the primary controller $\boldsymbol{u}_{\rm des}$ when available). The particular RTA used in this paper is designed for scenarios where the desired action $\boldsymbol{u}_{\rm des}$ is available to the RTA within rate and latency requirements. When the state is near a safety boundary, or $\boldsymbol{u}_{\rm des}$ will violate the safety boundary, the RTA modifies or substitutes a control signal (action) $\boldsymbol{u}_{\rm safe}$ that is actually sent to the environment (see Figure \ref{fig:RTARLDiagram}). If the state of the aircraft is safe and will stay safe under $\boldsymbol{u}_{\rm des}$, then the desired signal is passed to the plant ($\boldsymbol{u}_{\rm safe}=\boldsymbol{u}_{\rm des}$). RTA is designed to be completely independent of the primary control structure, such that the primary controller focuses on performance and mission objectives while the RTA focuses on safety assurance. RTA can be used during the training process to assure safe exploration, as well as after the NNCS is trained to assure safe operations \cite{dunlap2022run}.

RTA approaches could be categorized as filtering, switching, or latched \cite{dunlap2021comparing}. Filtering approaches make minor modifications to control signals to stay within safety limits. Switching approaches use a pre-planned safety maneuver to avoid an unsafe state before returning to the autonomous control. Filtering and switching approaches are often considered unlatched, where the system can frequently switch between the output of the primary controller and RTA. On the other hand, latched approaches switch to the output of the RTA for a longer period of time until specified criteria is met. 

One of the primary goals of this project is to conduct the first full scale aircraft flight test of a filtering RTA. One possible implementation of a filtering RTA is an optimization-based RTA, which uses gradient computations to create a set of barrier constraints, as presented in \cite{ames2016control,hobbs2021run, gurriet2018online,gurriet2020scalable,ames2019control}. By optimizing the actual input $\boldsymbol{u}_{\rm safe}$ to be as close to the desired input $\boldsymbol{u}_{\rm des}$ as possible, the optimization RTA is minimally invasive to the primary performance goals and results in smoother intervention when compared to a switching-based approach. In this case, the optimization filter uses a quadratic program, where the objective function is minimizing the two-norm difference between $\boldsymbol{u}_{\rm des}$ and $\boldsymbol{u}_{\rm safe}$. The filter construction is given by Eq. \ref{eq:optimization}.
\noindent \rule{1\columnwidth}{0.7pt}
\noindent \textbf{Optimization RTA}
\begin{equation}
\begin{split}
\boldsymbol{u}_{\text{act}}(\boldsymbol{x})={\text{argmin}} & \left\Vert \boldsymbol{u}_{\text{des}}-\boldsymbol{u}\right\Vert ^{2}\\
\text{s.t.} \quad & BC_i(\boldsymbol{x},\boldsymbol{u})\geq 0, \quad \forall i \in \{1,...,M\}
\end{split}\label{eq:optimization}
\end{equation}
\noindent \rule[7pt]{1\columnwidth}{0.7pt}
Here, $BC_i(\boldsymbol{x},\boldsymbol{u})$ represents a set of $M$ barrier constraints, where the constraints are satisfied when $BC_i(\boldsymbol{x},\boldsymbol{u}) \geq 0$.

In order to develop the barrier constraints, safety of the system must first be defined. To develop constraints for the system, the designer may first construct a set of $M$ inequality constraints $\varphi_i(\boldsymbol{x}): \mathcal{X}\to \mathbb{R}$,  $\forall i \in \{1,...,M\}$, where $\varphi_i(\boldsymbol{x}) \geq 0$ when each constraint is satisfied. Often for control systems, there exist states that adhere to all constraints at the current time but will eventually violate the constraints at a future time, due to limitations created by the admissible control set $\mathcal{U}$. For example, consider a wingman aircraft that is at the minimal safe separation distance from the lead aircraft, but with a velocity closing too fast to possibly rectify in time to avert a collision. Therefore, in order for the system to be safe, it must adhere to a set of $M$ control invariant inequality constraints $h_i(\boldsymbol{x}): \mathcal{X}\to \mathbb{R}$, $\forall i \in \{1,...,M\}$, where there must exist a control law $\boldsymbol{u}\in \mathcal{U}$ that renders all constraints forward invariant. Similarly, $h_i(\boldsymbol{x}) \geq 0$ when each constraint is satisfied. This set of constraints forms the \textit{safe set} $\mathcal{C}_{\rm S}$, which is defined as,
\begin{equation}
    \mathcal{C}_{\rm S} := \{\boldsymbol{x} \in \mathcal{X} \, | \, h_i(\boldsymbol{x}) \geq 0, \forall i \in \{1,...,M\} \}
\end{equation}
Safety can then be determined \textit{explicitly} offline, by simply verifying that $h_i(\boldsymbol{x})\geq0, \forall i \in \{1,...,M\}$ and therefore $\boldsymbol{x}\in \mathcal{C}_{\rm S}$.

To enforce safety, barrier constraints are then developed to enforce Nagumo's condition \cite{nagumo1942lage}. The purpose of these barrier constraints is to ensure that $\boldsymbol{x}$ will never leave $\mathcal{C}_{\rm S}$ by ensuring that for each constraint $\dot{h}_i(\boldsymbol{x})$ is never decreasing at any point along the boundary of $h_i(\boldsymbol{x})$. For each constraint, this condition is defined as,
\begin{equation}
    \dot{h}_i(\boldsymbol{x}) = \nabla h_i(\boldsymbol{x}) \dot{\boldsymbol{x}} = L_f h_i(\boldsymbol{x}) + L_g h_i (\boldsymbol{x}) \boldsymbol{u} \geq 0
\end{equation}
where $L_f$ and $L_g$ are Lie derivatives of $h_i$ along $f$ and $g$ respectively. To practically enforce this constraint, it is important that it is only applied near the boundary of $\mathcal{C}_{\rm S}$, and relaxed away from the boundary. To do this, a strengthening function $\alpha(x)$ is introduced that is a class-$\kappa$ function, is strictly increasing, and meets the condition $\alpha(0)=0$. The barrier constraint is then defined explicitly as,
\begin{equation}
    BC_i(\boldsymbol{x},\boldsymbol{u}) := L_f h_i(\boldsymbol{x}) + L_g h_i (\boldsymbol{x}) \boldsymbol{u} + \alpha(h_i(\boldsymbol{x}))
\end{equation}

%%%%%%%%%%%%%%%%%%%%% Problem Statement %%%%%%%%%%%%%%%%%%%%%%%%%%%%%%%%%%%%%%%%%%%%%
\section{Problem Statement} \label{Problem Statement}
\begin{figure} [htb]
\centering
\begin{subfigure}{.5\textwidth}
  \centering
  \includegraphics[width=.94\linewidth]{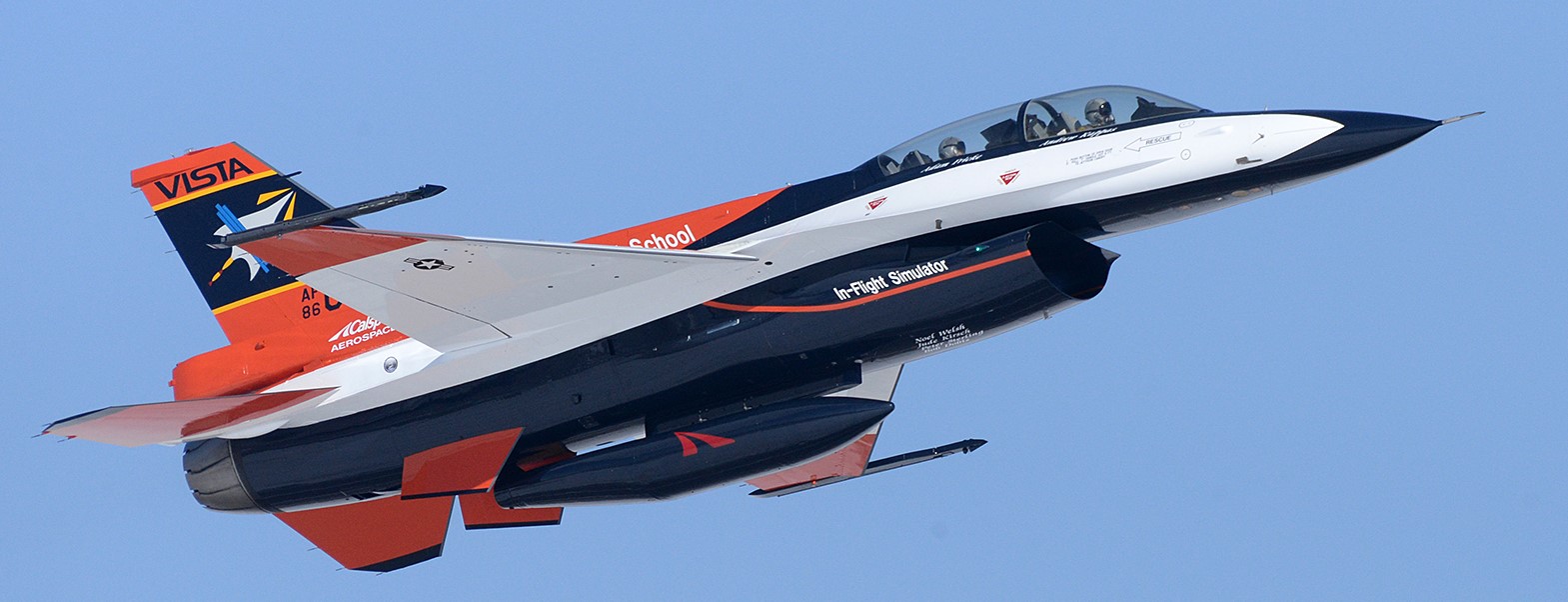}
  \caption{X-62 VISTA \cite{lloyd_2019}}
  \label{fig:sub1}
\end{subfigure}%
\begin{subfigure}{.5\textwidth}
  \centering
  \includegraphics[width=.9\linewidth]{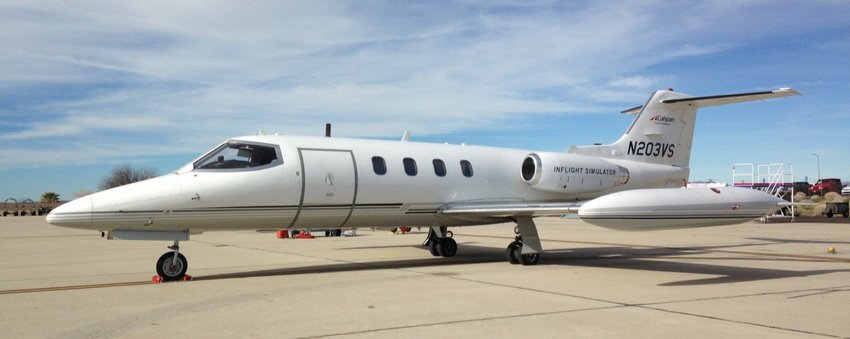}
  \caption{Calspan LJ-25 Learjet \cite{berger2017development}}
  \label{fig:sub2}
\end{subfigure}
\caption{Both the X-62 Variable Stability In-flight Simulator Test Aircraft (VISTA) and Calspan LJ-25 have the ability to simulate the dynamics of other aircraft and one use is to test theoretical flight control concepts.}
\label{fig:test}
\end{figure}

The goal of this project is to identify safety requirements for a hypothetical flight test called ``Have Leash,'' where an NNCS performs autonomous rejoin to formation flight while bounded by an RTA system providing geofence and collision avoidance. This project is scoped to safety analysis for a proof of concept flight test, which is a subset of what would be required for an operational system. The flight test safety analysis includes the following high-level special considerations.
\begin{itemize}
    \item A human safety pilot will be on board to serve as the last line of safety. 
    \item A special test aircraft will be used: either the General Dynamics X-62 VISTA experimental aircraft in Figure \ref{fig:sub1} or the Calspan LJ-25 Learjet in Figure \ref{fig:sub2}. The X-62 VISTA is owned and operated by the USAF at USAF TPS with the help of Calspan and Lockheed Martin ADP in support roles. It is derived from a Foreign Military Sales (FMS) F-16D from Lockheed Martin. The Learjet is owned and operated by Calspan, and is contracted for use at USAF TPS. 
    \item The X-62 VISTA features a VISTA Simulation System (VSS), while the Calspan Learjet features Variable Stability System (also VSS). Each VSS constrains experiments to a smaller flight operating envelope inside the full operating envelope of the Learjet or F-16 for safety. This gives buffer space for a safety pilot on either aircraft to recover the airplane should a test violate safety boundaries. Each VSS also has its own set of safety trips that turn off the experimental control system (such as the NNCS/RTA combo) or simulated aircraft dynamics and reverts to traditional F-16 or Learjet flying qualities. These VSS safety trips fall more generally into a category of EPM, and in this case acts as a second switching RTA.  
    \item The RTA will assure collision avoidance and geofencing using an filtering, optimization-based design.
    \item The NNCS will be trained using RL with SSCs (desirable but not required safety constraints). 
\end{itemize}

In addition to testing the NNCS and RTA on an aircraft within a restricted flight envelope, the following assumptions are made about the operating conditions of the aircraft: 
\begin{itemize}
    \item \textit{Flight Phase Scope}: The flight test will occur after a human pilot enters the flight test range and before the human pilot commands the aircraft to exit the flight test range. 
    \item \textit{Flight Test Coordination}: Each phase of all flight test events will be communicated with the ground control station.
    \item \textit{Live, Virtual, Constructive (LVC) Buildup}: The lead aircraft may be a (live) physical aircraft, a (virtual) manned simulator on the ground, or a (constructive) computer simulation of the live aircraft on the ground. The virtual and constructive options send substitute lead aircraft signals to the wingman, and are used to reduce the cost of flight test (paying for one aircraft in the air instead of two), as well as to reduce risk by testing components before two live aircraft are in the air and at risk of a potential collision.
    \item \textit{Scope}: The wingman aircraft is the primary entity responsible for collision avoidance with the lead, while the lead's responsibility is to predictably fly a pre-planned maneuver with bounded deviation. Collision avoidance is provided by the RTA design with a human safety pilot monitoring the wingman aircraft behavior, ready to take control at any time during the flight. 
    \item \textit{Time}: All flights will take place between dawn and dusk (during the day). No flights will take place at night.
    \item \textit{Weather}: Flights will only take place during visual meteorological conditions (VMC) required for visual flight rules (VFR).
    \item \textit{Traffic}: There is no other traffic close to the lead and wingman, with the exception of a photo chase aircraft following at a safe distance. In the event that another aircraft enters closely, the human safety pilot of the wingman test aircraft will assume control until the traffic has sufficient separation from the lead and wingman.
    \item \textit{Altitude}: The flight test will take place within a predetermined minimum and maximum altitude, which will prevent any terrain collisions.
    \item \textit{Dynamics}: The flight test system shall be able to assure safety of flight for reasonable lead aircraft dynamics (velocity, turn rates, altitude).
    \item \textit{Environment}: The flight test will be performed in an uncontested environment, with no adversarial interactions.
    \item \textit{Formation}: The flight test shall operate in a two-ship formation. 
    \item \textit{Outside Foreseeable Operating Conditions}: If any exception to foreseeable operating conditions occurs, the human pilots of the lead and wingman aircraft will take control and terminate the flight test.
    \item \textit{Detection of Conditions Outside Foreseeable Operations}: It is not anticipated that conditions outside foreseeable operating conditions will not be detected by the lead pilot, wingman pilot, or ground crew.
    \item \textit{Flight Test Plan}: Every flight scenario will be preplanned, and in some cases scripted.
    \item \textit{Flight Test Plan}: Specific tests focused on RTA (collision avoidance and geofence) will be completed with a scripted primary flight controller in addition to tests with than the NNCS.
    \item \textit{Flight Risk Reduction}: Every flight scenario will be evaluated in simulation prior to flight test.
\end{itemize}
\subsection{Concept of Operations}
The hypothetical flight test will include two aircraft: a manned flight lead aircraft and an surrogate unmanned wingman (test aircraft designed to fly autonomously, but with a human pilot on board who can take control at any time). The two aircraft will take off under the control of human pilots and fly to the flight test range. Once in the flight test range (inside the geofence and safely separated), the flight lead aircraft will fly pre-planned maneuvers. The wingman test aircraft will include at least two humans on board: a safety pilot and a test pilot/engineer. The safety pilot is responsible for flying the aircraft to the test range, controlling the aircraft between test points, maintaining situational awareness, monitoring for violations of SSCs, and being ready to take control of the aircraft at any point during test points. The test pilot/engineer selects specific test points and \textit{activates} (turns on) the NNCS/RTA system to perform the a test point. When the test point is complete, the test engineer \textit{deactivates} the NNCS/RTA system, and the safety pilot resumes control. The scope of this safety analysis is from the activation to the deactivation of the NNCS/RTA system. When the NNCS/RTA system is active, the EPM monitors for safety violations, or a signal from the safety pilot to switch control from NNCS/RTA to the safety pilot.

A set of test points is pre-defined by the flight test team. The control room (1) selects which test points the test engineer/pilot commands and (2) makes decisions for terminating tests. The Have Leash flight test is anticipated to achieve the following objectives (note that \textit{safely evaluate} means to effectively evaluate the system while maintaining safety, and LVC techniques will be utilized to reduce risk in a phased testing approach): 
%\textcolor{red}{What does safely evaluate mean?: It is \textbf{effectively} evaluating while maintaining safety...perform an effective flight test safely. Defining safety within the needed test.}
\begin{enumerate}
    \item Safely evaluate effectiveness of RTA providing safe separation (collision avoidance), validate simulation data.
    \item Safely evaluate effectiveness of RTA providing geofence (keep in zone), validate simulation data.
    \item Safely evaluate effectiveness of RTA providing geofence (keep in zone) and RTA providing safe separation (collision avoidance) simultaneously, validate simulation data. 
    \item Safely evaluate effectiveness of neural network control system performing rejoin when trained without  RTA, validate simulation data.
    \item Safely evaluate effectiveness of NNCS performing rejoin when trained with RTA, validate simulation data.
\end{enumerate}

\subsection{Rejoin to Formation Flight}

\subsubsection{Successful Rejoin Point Definition}
\begin{figure}
     \centering
     \begin{subfigure}[b]{0.49\textwidth}
         \centering
         \includegraphics[width=\textwidth]{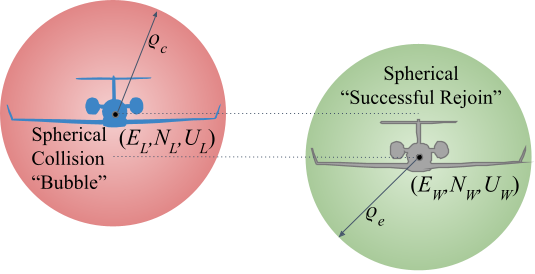}
         \caption{Safe rejoin as pictured from behind the lead and wingman.} 
         \label{fig:SafeRejoinSpheres_a}
     \end{subfigure}
     %\hfill
     \begin{subfigure}[b]{0.49\textwidth}
         \centering
         \includegraphics[width=\textwidth]{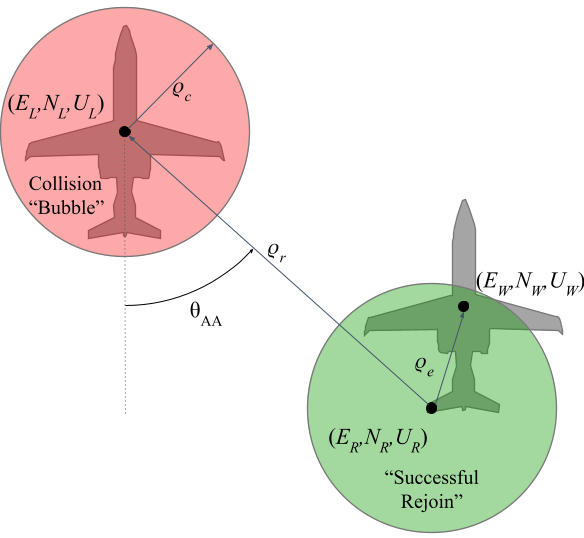}
         \caption{Safe rejoin as pictured from above the lead and wingman.}
         \label{fig:SafeRejoinSpheres_b}
     \end{subfigure}
        \caption{Depictions of safe rejoin.}
        \label{fig:saferejoin}
\end{figure}

The objective of the NNCS is to conduct a rejoin, e.g. move the wingman aircraft to a position defined by a relative lateral distance $\varrho_r$, and an aspect angle $\theta_{AA}$ measured from vector pointing through the back of the lead aircraft to the vector pointing to the wingman's location. While a rejoin point is one convenient way to define success, another approach is to identify the ideal speed for station keeping based on $\theta_{AA}$ and range for inside and outside turns. Note that the total range of possible rejoin positions is limited by a constrained flight envelope monitored by the EPM.

The aspect angle $\theta_{AA}$ and distance $\varrho_r$ correspond to a formation flight position that a human flight lead would command a wingman to enter.
This relative position command can be converted to Cartesian coordinates in a East-North-Up (ENU) reference frame using aircraft headings $\psi$ by Eq. \ref{e:RejoinPosition}.
\begin{equation}
 \label{e:RejoinPosition}
 \begin{split}
     E_R & =  E_L+\varrho_r\cos(\psi_L+\pi+\theta_{AA}) \\ 
     N_R & =  N_L+\varrho_r\sin(\psi_L+\pi+\theta_{AA}) \\
 \end{split}
\end{equation}
Note that heading (direction that the aircraft nose is pointed) is measured clockwise from North. The rejoin point $(E_R, N_R, U_R)$ is defined as a moving point relative to the wingman via aspect angle $\theta_{AA}$, and rejoin distance in the E-N frame $\varrho_r$. The position of the lead and wingman are $(E_L, N_L, U_L)$ and $(E_W, N_W, U_W)$, respectively.
A rejoin is successful if the duration that the wingman is within an acceptable range for the rejoin point $t_{rejoin}$ is greater than a minimum amount of time $t_{success}$. 
\begin{equation}
 \label{e:HelloRejoinSuccess1}
    \gamma_{rejoin}: \Big( \sqrt{(E_W - E_R)^2 + (N_W-N_R)^2+ (U_W-U_R)^2} \leq  \varrho_e \Big) \land (t_{rejoin} \geq t_{success})
\end{equation}
%
%\begin{figure}[htb!]% order of placement preference: here, top, bottom
%\centering
%\includegraphics[width =0.5\textwidth]{Figures/GeneralRejoin2D.png}
%\caption{Depiction of general rejoin task}
%\label{f:GeneralRejoin2D}
%\end{figure}
%
\subsubsection{Successful Rejoin Speed and Bank Angle Definition}
In addition to the point-based rejoin definition, a successful rejoin can be defined by a target bank $\phi_r$ and speed $v_r$ computed from the lead's speed $v_L$ in knots true air speed, and lead's bank angle $\phi_L$ in degrees. Intermediate computations include the lead turn radius $R_L$ in feet and the lead turn rate $\omega_L$ in deg/s.

\begin{algorithm}
\caption{Computing target bank angle and speed}\label{alg:bankspeed}
\begin{algorithmic}
\If{$\phi_L$ == 0}
    \State $\phi_r \gets \phi_L$
    \State $v_r \gets v_L$  
\Else
    \State $R_L \gets (v_L^2)/(11.26\tan\phi_L)$
    \State $\omega_L \gets (1091\tan\phi_L)(v_L) $
    \If{$(\phi_L <0 \land \theta_{AA}<0)\lor (\phi_L>0\land \theta_{AA}>0$)}
        \State $v_r \gets (11.26\omega_L(R_L-\varrho_r\sin\theta_{AA}))/1091$
    \Else 
        \State $v_r \gets (11.26\omega_L(R_L+\varrho_r\sin\theta_{AA}))/1091$
    \EndIf
    \State $\phi_r \gets \arctan(v_r\omega_r/1091)$
\EndIf
\end{algorithmic}
\end{algorithm}

\subsection{Run Time Assurance}

This problem presents a system with three RTAs that operate asynchronously on a synced input, with one RTA (the safety pilot) that is nondeterministic. Each RTA has different monitoring responsibilities. The human safety pilot monitors SSCs and provides a backup to the collision avoidance and geofence RTAs. The EPM only monitors for limit violations of specific aircraft state variables like angle of attack $\alpha$, sideslip angle $\beta$, and velocity. The RTA referred to as ``RTA'' in this paper provides collision avoidance and geofence. The RTA must be designed within the safety thresholds of the EPM, and the human pilot (collision avoidance and geofence only).

\subsubsection{Collision Avoidance}
The objective of the collision avoidance is to assure: 
\begin{equation}
 \label{e:collisionavoidance}
    \varphi_{collision avoidance}: \sqrt{(E_W - E_L)^2 + (N_W-N_L)^2+ (U_W-U_L)^2} - \varrho_c 
\end{equation}
where $\varrho_c$ is the minimum safe separation distance from the collision zone, and $\varphi_{collision avoidance} \geq 0$ when the safety constraint is satisfied.

\subsubsection{Geofence}
\begin{figure}[hbt!]
\centering
\includegraphics[width=0.75\columnwidth]{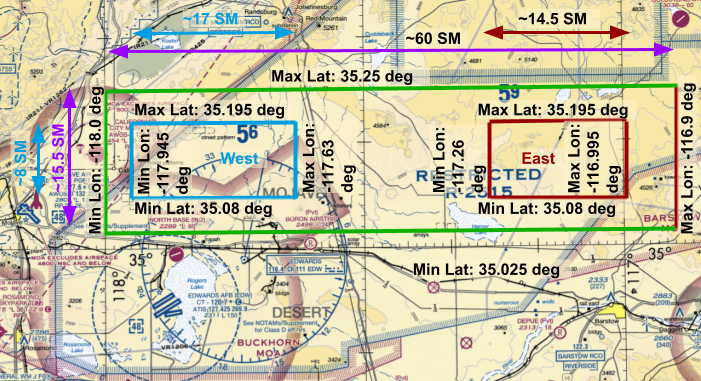}
\caption{Example flight test range geofences.}
\label{fig:Geofence}
\end{figure}

Generally geofences are keep-in or keep-out zones that take the form of a) polygons where the latitude and longitude of each vertex is defined, b) max and min latitude and longitude rectangles, or c) a point and a distance. But they may also be defined by specific types of airspace. 
There are several types of airspace over the Edwards Air Force Base flight testing area, pictured in Figure \ref{fig:Geofence}. The full East-West Range is marked in green, a subset called the “West” area is bordered in blue, and another subset called the “East” area is bordered in red.\footnote{A full sectional chart of the area is available at: \url{https://aeronav.faa.gov/visual/10-07-2021/PDFs/Los_Angeles.pdf}}. 

\subsubsection{Envelope Protection Monitor}
In the case of both the VISTA and Calspan LJ-25 Learjet, the EPM is provided by the VSS which switches control from the system under test (NNCS/RTA) to the safety pilot when a variety of limits are violated.  The VSS safety trips and reversion to a human safety pilot are a form of RTA. The EPM is separate from the collision avoidance and geofence RTAs, which will be discussed more in Section \ref{sec:STAMPModel}. The VSS is the core capability that enables in-flight testing of novel control algorithms. To assure safety of flight, the VSS operates in a constrained flight envelope of the Learjet or VISTA aircraft. If the novel control algorithm causes the aircraft to go outside the constrained VSS envelope, the simulator is turned off and control is reverted to the Learjet's or VISTA's actual flight dynamics controlled by a safety pilot. The VSS on the Learjet and VISTA have different safety constraints, and example variables that may be monitored by the VSS include:
\begin{multicols}{2}
\begin{itemize}
    \item $\delta_e$, elevator deflection, deg,  
    \item $\delta_a$, aileron deflection, deg, 
    \item $\delta_r$, rudder deflection, deg,     
    \item $\dot{\delta}_e$, elevator deflection rate, deg/s,  
    \item $\dot{\delta_a}$, aileron deflection rate, deg/s, 
    \item $\dot{\delta_r}$, rudder deflection rate, deg/s, 
    \item $\alpha$, angle of attack (AOA), deg,
    \item $\beta$, sideslip angle, deg,
    \item $N_z$, normal load factor,g,
    \item $N_y$, side load factor, g,
    \item $V_c$, calibrated airspeed, knots,
    \item $\phi$, roll, deg,
    \item $\dot{\phi}$, roll rate, deg/s,
    \item $\psi$, yaw, deg,
    \item $r_t$, horizontal tail rate limit, deg/s, and
    \item surface structural limits.
\end{itemize}
\end{multicols}
Additional details can be found in \cite{ackerman2017evaluation,CallaghanThesis}. For this research, the safety trips can be considered done conditions (termination criteria for a training episode) for the RL training and the required operating envelope for the RTA system. %The safety trips for the VSS are described in Table \ref{tab:VSSLimits}. 

\section{Results} \label{Results}
In this section, specific results for Have Leash are presented. First, accidents, hazards, and safety constraints are listed before introducing new concepts of SSCs and CCSs. Next, a functional control system block diagram is created for the Have Leash flight test system, and analyzed with STPA to find UCAs, scenarios, and requirements.

\subsection{Accidents, Hazards, and Safety Constraints}
The following accidents were identified in line with the STAMP/STPA methodology: 
\begin{itemize}%\setlength\itemsep{-.3em}
    \item[{[A1]}] Death or injury of a person.\\
    \textbf{Rationale}: Human injury or death during flight test is unacceptable.
    \item[{[A2]}] Ground property is damaged.\\
    \textbf{Rationale}: Flight test mishaps may result in ground property damage. Flight tests are generally conducted in remote areas away from people to prevent loss of ground property and human life (covered by A1).
    \item[{[A3]}] Aircraft is damaged or destroyed.\\
    \textbf{Rationale}: The test aircraft envisioned for this research are one-of-a-kind national assets with exorbitant replacement costs, and the loss would also result in the lost opportunity to test other new technologies.
    \item[{[A4]}] Aircraft faces consequences of entering prohibited airspace.\\
    \textbf{Rationale}: Leaving the flight test range introduces hazards to private citizens and other entities operating in the national airspace, while entering prohibited airspace in operational contexts might signal unintended aggression and elicit an undesirable response from other nations or organizations.
    \item[{[A5]}] Mission goals are not achieved.\\
    \textbf{Rationale}: Not achieving mission goals not only represents an unacceptable loss (economic loss of the cost of the test/mission and loss of data from test/mission), but also represents the need for another flight to collect test data or finish the mission, which adds additional risk to the human pilots who have to re-fly the mission or test.
\end{itemize}
%\vspace{5mm}
%
\begin{table}[htb!]
\begin{center}
\caption{\label{t:AHSCI} Hazards and Safety Constraints}
%\centering
%\setlength\extrarowheight{-10pt}
\begin{tabular}{cccccll}
[A1]       & [A2]       &[A3]        & [A4]       &[A5]        & Hazard                         & Safety Constraint \\\hline
\checkmark & \checkmark & \checkmark &            &            & [H1] Aircraft violates minimum & [C1] Aircraft shall satisfy minimum \\
           &            &            &            &            & separation from other aircraft & separation requirements from  \\
           &            &            &            &            & or terrain.                    & other aircraft and objects.\\\hline
\checkmark & \checkmark & \checkmark &            &            & [H2] Aircraft loses control.  & [C2] Aircraft shall maintain control.\\\hline
\checkmark & \checkmark & \checkmark & \checkmark &            & [H3] Aircraft exits allowable  & [C3] Aircraft shall not depart \\
           &            &            &            &            & airspace.                      &the allowable airspace.\\\hline
\checkmark & \checkmark & \checkmark &            &            & [H4] Aircraft conducts an      & [C4] Aircraft shall not conduct \\
           &            &            &            &            & excessively aggressive maneuver &an excessively aggressive \\
           &            &            &            &            & (that causes harm to safety    & maneuver (stay within pilot,\\
           &            &            &            &            & pilot or violates aircraft     & structural limits).\\
           &            &            &            &            & or component structural limits). &\\\hline
           &            &            &            & \checkmark & [H5] Aircraft does not keep    & [C5] Aircraft shall properly store \\
           &            &            &            &            & data until it can be offloaded & data until it is offloaded for \\
           &            &            &            &            & for analysis.                  & analysis.\\\hline
           &            &            &            & \checkmark & [H6] Aircraft does not execute & [C6] Aircraft shall execute planned \\
           &            &            &            &            & planned operations.            & operations. \\\hline
\end{tabular}
\end{center}
\end{table}
Constructing the hazards in this project involved analysis of previous MUMT work as well as consideration of this particular challenge. Hazards 1-3 were inspired by previous MUMT STAMP/STPA research \cite{montes2016using,hobbs2018early,robertson2019systems}. %Hazard 1 may be found across all three previous MUMT STAMP/STPA projects was inspired by the common hazard across all three manned-unmanned teaming projects, whether it was defined as violating ''minimum separation distance to other flying objects'' \cite{montes2016using,hobbs2018early}, ''terrain closure limits,” \cite{montes2016using,hobbs2018early}, or ``minimum separation from other aircraft or terrain'' \cite{robertson2019systems}.
%Likewise, Hazard 2 was inspired by all three MUMT sources in forms of ``Aircraft departs aerodynamically stable flight,” \cite{montes2016using} ``an aircraft departs stable flight'' \cite{hobbs2018early}, and ``aircraft loss of control (includes departure from stable flight)” \cite{robertson2019systems}. Hazard 3 was inspired by ``Aircraft exits allowable testing area” \cite{montes2016using},  ``aircraft departs approved airspace,” \cite{hobbs2018early, robertson2019systems}. Hazard 6 is inspired by \cite{robertson2019systems} and represents a path for UCAs in the category of coordination and communication failures. Some hazards from previous work were not applicable to this due to the difference in the target mission, where this project focuses on safe formation flight.
%
%Hazards 4 and 5 are unique to this project, but inspired by previous work. 
Hazard 4 was inspired by “do no harm” requirements in development of the Automatic Ground Collision Avoidance System (Auto GCAS), specifically that the ``the automatic recovery shall not cause harm to pilot, aircraft, or components'' \cite{hobbs2020elicitation}.  Hazards 5 and 6 are new hazards to this research, because data capture is the mission of the flight test, and often the mission of operational UAVs, so the largest contributor to Accident 5 (loss of mission) is loss of data, or never capturing it in the first place.

\subsection{Soft Safety Constraints}
During the course of the research, a need was identified for the specification of SSCs, which are desirable conditions for the system under test, but are not strictly enforced. In many cases, constraints like maintaining line of sight between the aircraft or staying out of the jet wash of the lead, are things that introduce higher risk the longer they are violated. For instance, depending on the test aircraft, if a maneuver passes quickly through the jet wash it may not be an issue, but choosing to loiter there can lead to a excessive wear and tear on aircraft, loss of control, and degraded situational awareness for the lead aircraft. Additionally, loosing line of sight for a few seconds can be safe under some circumstances. In practice, these are used to train the RL agent to choose behavior that would be safer for flight test. Considerations like these were captured over the course of the project, and are listed below. 
\begin{itemize}[leftmargin=0.5in]
    \item [{[SSC1]}] \textbf{\textit{Tail Nose Separation Criteria (TNSC)}}: When the Altitude Independent Slant Range (AISL) is less than 500 feet, the wingman aircraft’s nose should be at least 50 feet behind the tail of the lead aircraft, and the aspect angle shall be less than +/- 90 degrees (i.e. behind the 3-9 line). The formal definition of this criteria is defined by several variables. 
    \begin{figure}[htb!]
    \centering
    \includegraphics[width=.45\textwidth]{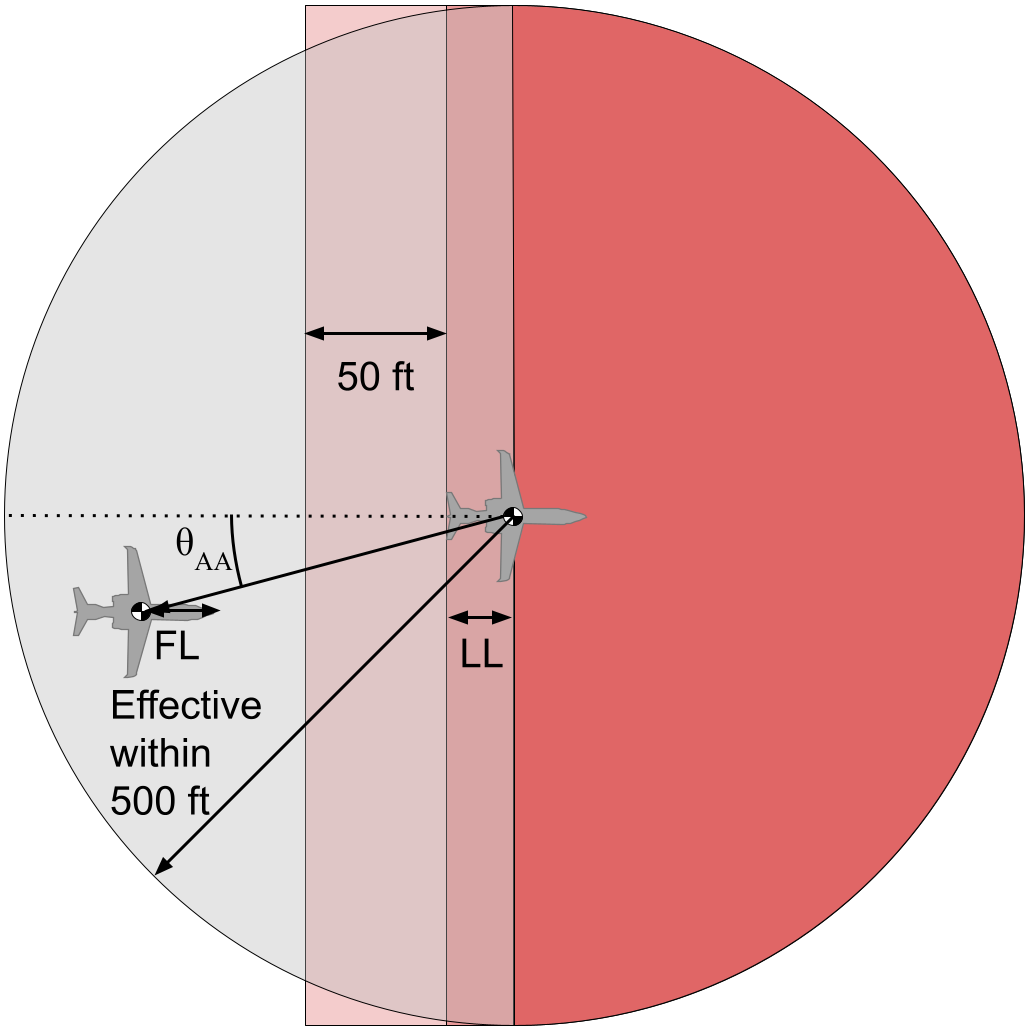}
    \caption{Tail Nose Separation Criteria (TNSC). \label{fig:TailNoseSeparation}}
    \end{figure}
    The Follow Length (FL) is the length from the nose of the wingman aircraft to the wingman position reference point. For the Learjet and F-16, a FL of 25 ft is used. Note that the total length of the VISTA is 48 feet 7 inches. The Lead Length (LL) is the length from the tail of the lead aircraft to the lead aircraft reference point. The aspect angle $\theta_{AA}$ is the angle between the tail of the lead aircraft and the relative position vector of the wingman aircraft. To be behind the aircraft, the absolute value of the $\theta_{AA}$ must be less than 90 degrees. The AISL is computed from the position reports of the lead and wingman aircraft, using the East  and North positions of both the lead and wingman position in a ENU reference frame, as follows: 
    \begin{equation}
        AISL = \sqrt{(E_L-E_W)^2+(N_L-N_W)^2}.
    \end{equation}
    The Tail Nose Separation (TNS) is computed from
    \begin{equation}
        TNS = (AISL)(\cos(|\theta_{AA}|))-LL-FL. 
    \end{equation}
    The TNSC SSC1 property can then be formalized as:
    \begin{equation}
        \gamma_{SSC1}: AISL < 500 \implies (TNS>TNSC)\land(|\theta_{AA}|<90).
    \end{equation}
%\begin{figure}[htb!]
%\centering
%\includegraphics[width=.5\textwidth]{Figures/AltitudeSeparation2.png}
%\caption{Altitude Follow Stack Criteria (AFSC). \label{fig:AltitudeSeparation2}}
%\end{figure}
\begin{figure} [htb]
     \centering
     \begin{subfigure}[b]{0.55\textwidth}
         \centering
         \includegraphics[width=\textwidth]{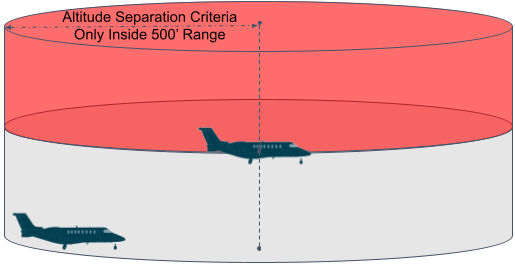}
        \caption{Altitude Follow Stack Criteria (AFSC). \label{fig:AltitudeSeparation2}}
     \end{subfigure}
     \hfill
     \begin{subfigure}[b]{0.25\textwidth}
        \centering
        \includegraphics[width=\textwidth]{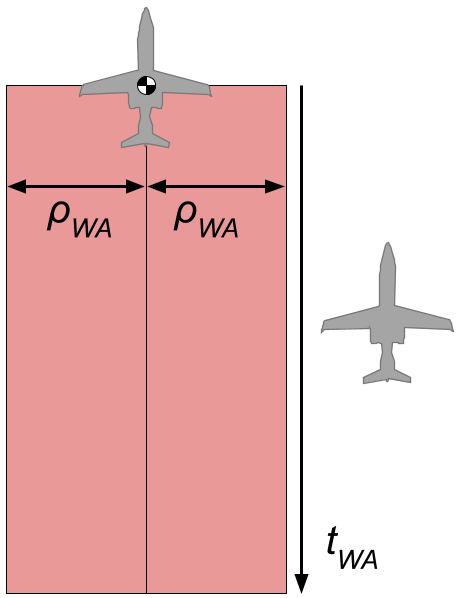}
        \caption{Jet Wash Exclusion Zone Criteria (JEC). \label{fig:WakeAvoidance}}
     \end{subfigure}
     \hfill
        \caption{Soft Safety Constraint}
        \label{fig:three graphs}
\end{figure}
    \item [{[SSC2]}] \textbf{\textit{Altitude Follow Stack Criteria (AFSC)}}: For the safety pilot to maintain visual contact with the lead aircraft, an AFSC is used. In the Learjet, additional consideration will be needed for which side the safety pilot is on (usually left seat) and the current maneuver the pilot is doing. In the VISTA aircraft, the safety pilot will be in the front seat while the test engineer is in the backseat. When the AISL is less than 500 feet, the wingman aircraft nose should be below be at or below the lead aircraft altitude to allow safety pilot to maintain visual contact during final stages of the rejoin closure and alignment and when transitioning to station keeping. In an ENU reference frame, this is formalized as:
    \begin{equation}
        \gamma_{SSC2}: AISL < 500 \implies (U_L>U_W).
    \end{equation}
    \item [{[SSC3]}] \textbf{\textit{Smooth Flight Criteria (SFC)}}: When the AISL is between 100-300 feet, the wingman should minimize abrupt high g-load turns, accelerations, and climbs in towards the lead aircraft (i.e. maintain a constant closure of less than 10 ft/s). Formally,
    \begin{equation}
        \gamma_{SSC3}: AISL\in{100,300} \implies \sqrt{(\dot{E}_L-\dot{E}_W)^2+(\dot{N}_L-\dot{N}_W)^2}.
    \end{equation}
    \item [{[SSC4]}] \textbf{\textit{Jet Wash Exclusion Zone Criteria (JEC)}}: Wingman maintains a 60-second line segment of lead aircraft LLA positions creating a flight exclusion zone of 50 feet from line segment. Note that this is an oversimplification of jetwash and higher fidelity models may be incorporated at a later date.
%
%\begin{figure}[htb!]
%\centering
%\includegraphics[width=.25\textwidth]{Figures/WakeAvoidance.png}
%\caption{Jet Wash Exclusion Zone Criteria (JEC). \label{fig:WakeAvoidance}}
%\end{figure}
%

\end{itemize}

\subsection{Common Cause Scenarios}
During stakeholder interviews, several high level CCSs were identified. Because it was sometimes difficult to determine whether these constituted hazards or scenarios, the general metric used was “could this cause something already on the hazard list?” In STAMP/STPA, causal scenario generation results from the process of analyzing UCAs. A formal definition of “causal scenarios” is not provided in traditional STAMP/STPA literature. So here a new “Common Causal Scenario” is defined as sets of conditions identified prior to system level analysis that could lead to a hazard, which may include prior lists of common failures from documentation or subject matter expertise. This section captures some initial CCSs prior to the STPA analysis, followed by the hazards they contribute to. 

\begin{itemize}[leftmargin=0.6in]
    \item [{[CCS1]}] An aircraft is on a collision course with another aircraft. {[H1]}
    \item [{[CCS2]}] An aircraft fails to detect other aircraft on collision course. {[H1]}
    \item [{[CCS3]}] The wingman aircraft maneuvers in a manner unexpected by flight lead. {[H1, H3, H4]}
    \item [{[CCS4]}] The wingman aircraft forces trajectory of lead. {[H1, H2, H3, H4, H6]}
    \item [{[CCS5]}] An aircraft flies through jet wash of another aircraft. {[H2]}
    \item [{[CCS6]}] An aircraft experiences environmental conditions outside safe ranges (e.g. icing). {[H2]}
    \item [{[CCS7]}] An aircraft experiences GPS dropout. {[H1, H3, H6]}
    \item [{[CCS8]}] An aircraft experiences communication/data dropout. {[H1, H3, H5, H6]}
    \item [{[CCS9]}] An aircraft loses or has degraded communication. {[H1, H3, H5, H6]}
    \item [{[CCS10]}] An aircraft experiences a loss of sensing. {[H1, H3, H6]}
    \item [{[CCS11]}] An aircraft departs flight/contingency plan. {[H1, H3, H6]}
    \item [{[CCS12]}] A pilot’s full capabilities are compromised in some way (e.g. a pilot is distracted by autonomous flight behaviors, suffers excessive workload / task saturation, experiences target fixation, experiences gravity-induced loss of consciousness (GLOC), or is spatially disoriented). {[H1, H2, H3, H4, H6]}
    \item [{[CCS13]}] A malicious actor gains control of aircraft. {[H1, H2, H3, H4 H5, H6]}
    \item [{[CCS14]}] A malicious actor gains access to data. {[H6]}
\end{itemize}

\subsection{Systems Theoretic Accident Model and Processing Block Diagram} \label{sec:STAMPModel}
\begin{figure}[htb!]
\centering
\includegraphics[width=.55\textwidth]{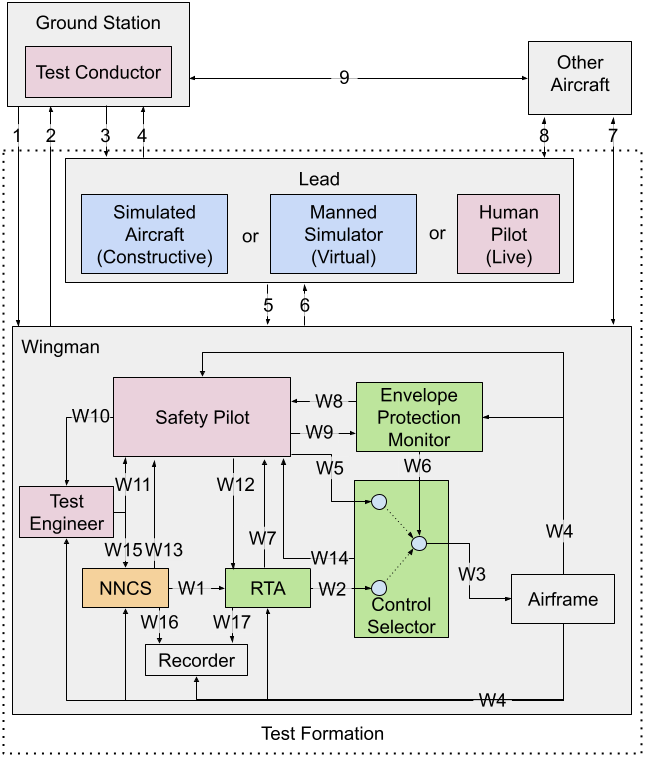}
\caption{STAMP control structure block diagram for RTA-wrapped NNCS flight tests. The gray boxes describe Level 1 elements, while level 2 elements include humans in pink, ground-based simulators in blue, computer safety systems in green, and the high-risk/high-performance elements in orange. \label{fig:STAMP_FlightTest2}}
\end{figure}

The STAMP control structure block diagram in this research contains two levels, with Level 1 elements in gray containing Level 2 elements, as shown in Figure \ref{fig:STAMP_FlightTest2}.
The Level 1 STAMP Model includes 4 high level components: the ground station, the wingman aircraft, the lead, and other aircraft. The communication between the components is summarized in Table \ref{t:STAMP1}. This STAMP control structure block diagram is arranged semi-hierarchically so that systems with lower authority are placed below systems with higher authority. Control information is generally sent from higher level systems to lower level systems, which return feedback to close a loop. Sometimes, having two systems at the same level of the hierarchy can highlight a design flaw that does not provide guidance on how to reconcile conflicting control information. In practice, the ground station, wingman, or lead, all have the authority to stop flight testing at any time due to safety of flight concerns, so they are placed slightly offset to represent priorities simultaneously with their parallel safety of flight roles. 
The ground station includes the test conductor, who is a human supervising the mission and can see telemetry from the aircraft. The test conductor can also speak with the pilot of the wingman aircraft, lead aircraft and pilots of other aircraft in the area.  The ground station is in control of the overall mission, and coordinates with external groups including the manager of the airspace and all missions flown within it. Since this research focuses mostly on the NNCS and RTA, the agencies that the ground station interacts with are abstracted out, but could be included in future work. For an example of how these might be integrated, one can reference \cite{montes2016using}. The lead aircraft is placed below the ground station in the diagram to be representative of a transition to operations where the lead would be in command of the wingman. The lead can stop the testing due to a safety of flight concern; however, in the flight test the lead’s task is simple and the lead pilot may not always have line of sight to the wingman. The wingman aircraft is placed lowest, to be representative of operations; however, the wingman safety pilot in the flight test should have line of sight the lead at all times and has final say in the safety of flight. 

In flight test, the wingman aircraft is an “unmanned surrogate” aircraft that is intended to simulate a flight test of an unmanned aircraft with a human safety pilot on board. In addition to the human pilot, the aircraft may include crew such as an additional safety pilot and/or flight test engineer(s). The wingman aircraft is responsible for testing the safe rejoin maneuver under the control of an NNCS that is wrapped by a collision avoidance and geofence RTA.

The other aircraft could be a chase plane taking video or photographs of the flight test, other aircraft performing unrelated flight tests, or other traffic in the area. If used, the chase aircraft is expected to be in radio contact and have line of sight to the lead and wingman. Other unrelated flights may or may not have radio contact or line of sight. In addition, it is possible for general aviation aircraft practicing Visual Flight Rules (VFR) to be in the airspace, and it is not anticipated that these aircraft will be in communication with the flight test aircraft or ground station. The East-West region over Edwards Air Force Base where the flight test is expected to take place contains a mix of Military Operating Areas, where general aviation aircraft are allowed and advised to exercise extreme caution, and Restricted Areas, where general aviation aircraft can fly with permission of the controlling agency.

\begin{table}[htb!]
\begin{center}
\caption{\label{t:STAMP1} Level 1 STAMP Signals\\(GS=Ground Station, W=Wingman, L=Lead, OA=Other Aircraft)}
%\centering
%\setlength\extrarowheight{-10pt}
\begin{tabular}{cccl}
	&	From	&	To	&	Information	\\\hline
1	&	GS	&	W	&	Wingman Control Input Confirmation	\\
	&		&		&	Wingman Position Report Confirmation	\\
	&		&		&	Voice Commands/Coordination/Safety Advisories	\\\hline
2	&	W	&	GS	&	Wingman Control Input Report	\\
	&		&		&	Wingman Position Report	\\
	&		&		&	Voice Coordination / Safety Concerns	\\\hline
3	&	GS	&	L	&	Entity Report Confirmation	\\
	&		&		&	Aircraft Control Request Status	\\
	&		&		&	Voice Coordination / Safety Concerns	\\\hline
4	&	L	&	GS	&	Lead Position Report	\\
	&		&		&	Voice Coordination / Safety Concerns	\\\hline
5	&	L	&	W	&	Lead Position Report	\\
	&		&		&	Voice Coordination / Safety Concerns	\\\hline
6	&	W	&	L	&	Wingman Position Report	\\
	&		&		&	Voice Coordination / Safety Concerns	\\\hline
7	&	W/OA	&	OA/W	&	Voice Coordination / Safety Concerns (Possible)	\\
	&		&		&	Position Report/Entity Report (Possible)	\\\hline
8	&	L/OA	&	OA/L	&	Voice Coordination / Safety Concerns (Possible)	\\
	&		&		&	Entity Report (Possible)	\\\hline
9	&	GS/OA	&	OA/GS	&	Voice Coordination / Safety Concerns (Possible)	\\
	&		&		&	Entity Report (Possible)	\\\hline
\end{tabular}
\end{center}
\end{table}

In the Level 2 Wingman STAMP Model, the airframe represents all components of the aircraft outside of the control signal (to include actuators and sensors). The airframe sends a wingman position report, lead position reports, and applicable entity reports to the NNCS, RTA, Human Pilot, Crew, EPM, and data Recorder (W4). A position report contains state information about the wingman or lead, and entity reports contain state information about other aircraft.
\begin{table}[htb!]
\begin{center}
\captionsetup{justification=centering}
\caption{Level 2 Wingman STAMP Signals\\(NNCS = Neural Network Control System, RTA = Run Time Assurance, CS = Control Selector, EPM = Envelope Protection Monitor, VSS = CS+EPM = Variable Stability System)}
\label{t:Lvl2sig} 
%\centering
%\setlength\extrarowheight{-10pt}
\begin{tabular}{cccl}
	&	Source	&	Sink	&	Information	\\\hline
W1	&	NNCS	&	RTA	&	Pitch, Roll, Speed, Yaw Command (desired)	\\\hline
W2	&	RTA	&	CS	&	Pitch, Roll, Speed, Yaw Command (filtered from NNCS); Errors	\\\hline
W3	&	CS	&	Airframe	&	Pitch, Roll, Speed, Yaw Command (filtered or pilot-provided)	\\\hline
W4	&	Airframe	&	RTA, NNCS, Crew,	&	Lead Position Report, Wingman Position Report, Entity Reports	\\
& & Recorder, Pilot, EPM & \\\hline
W5	&	Pilot	&	CS	&	Pitch, Roll, Speed, Yaw Command (pilot-provided)	\\\hline
W6	&	EPM	&	CS	&	Control set to come from Pilot or NNCS→RTA	\\\hline
W7	&	RTA	&	Pilot/Engineer	&	Status (RTA Active, RTA Intervening, RTA Off, RTA faults, 	\\
 & & &  RTA data dropout)\\\hline
W8	&	EPM	&	Pilot/Engineer	&	Status (Pilot or NNCS→RTA control; monitoring active/faults)	\\\hline
W9	&	Pilot	&	EPM	&	Switch to Pilot control	\\\hline
W10	&	Pilot	&	Engineer	&	Voice Coordination / Safety Concerns (Possible)	\\\hline
W11	&	Engineer&	Pilot	&	Voice Coordination / Safety Concerns (Possible)	\\\hline
W12	&	Pilot/Engineer	&	RTA	    &	Risk level, on/off, etc.	\\\hline
W13	&	NNCS	&	Pilot/Engineer	&	Current Status / explanation of plan, presence of faults 	\\\hline
W14	&	CS    &	Pilot/Engineer	&	CS selection state, when no signal received in W5 or W2	\\\hline
W15 &   Pilot/Engineer    &	NNCS	& NNCS settings \\\hline
W16	&	NNCS	    &	Recorder	&	NNCS inputs and outputs	\\\hline
W17	&	RTA	    &	Recorder	&	RTA inputs, outputs, and active/inactive state	\\\hline
\end{tabular}
\end{center}
\end{table}

A control signal may come from multiple sources and contains pitch, roll, speed, and yaw commands. These commands can be at different mutually-exclusive innerloop, outerloop, navigation, or guidance levels. Along the bottom of the diagram, the NNCS sends a desired control signal to the RTA module (W1). The RTA receives the desired control signal and outputs a filtered control signal to the CS (W2): if all elements of the desired control signal satisfy geofence and collision avoidance constraints, the filtered control signal is equal to the desired control signal; otherwise the filtered control signal is the closest safe output to the desired control signal. If there is a fault in the RTA that doesn't allow it to provide protection, a fault signal is sent to the CS through (W2). The human safety pilot and engineer can see the status of the RTA (W7, W7$\to$ W11), and can turn the RTA off or adjust RTA parameters (W12). The human safety pilot and engineer are also able to see status and data from the NNCS (W13, W13 $\to$ W10)

In parallel to the NNCS and RTA, a human safety pilot is monitoring the state of the aircraft and is ready to send pilot-provided control signal at any time (W5). The human pilot also receives the state of the RTA (whether it is active, intervening, or off) (W7) and the EPM (monitoring active, or presence of monitoring faults, and whether filtered or pilot-provided signal is applied) (W8). If the pilot desires to take control, he or she can send a switch command as well as desired signals to the EPM (W9). The EPM will decide to send a signal to the control selector (CS) via (W6) to switch to the pilot-provided control (W5) signal if a) predefined aircraft EPM safety constraints are violated in the position report, or b) the human pilot sends a command through the EPM to take control of the aircraft (via W9); otherwise the filtered control signal will be allowed (W2). This switch command from the EPM (W6) is sent to the CS which sends either the filtered or pilot-provided control signal to the airframe (W3). Then the loop repeats at a specified frequency. Data about the NNCS and RTA are sent to the data Recorder for post-processing analysis via (W16) and (W17).

Crew members on board may communicate safety concerns and other coordination to (W11) or from (W10) the pilot. Note that a human-machine interface (HMI) is excluded from Figure \ref{fig:STAMP_FlightTest2} to reduce complexity. The Flight Test engineer also has access to W8 and W9 through W10 and W11. The HMI is used to activate and deactivate the NNCS, view status of subsystems, and command changes such as updated geofence or collision avoidance requirements. The test engineer or safety pilot may input settings on the NNCS via (W16).

\subsection{Systems Theoretic Process Analysis Results}

The process of going through the STPA analysis helped to highlight the need for additional feedback signals (W12-W17), and added richness to the functions of each controller. For example, the safety pilot or engineer will need to enter geofence specifications on the fly during the flight test as range conditions change. It was identified that a validity check was required to ensure the wingman aircraft started far enough away from the geofence that preventing violations was possible as well as far enough away from the lead aircraft that safety constraints were satisfied during initialization of the NNCS/RTA test point. After debate on whether this check occurs at the RTA or at the interface (before traveling down W12), it was determined that the check and feedback to the pilot should happen before being transmitted to the RTA. The STPA process also \textit{highlighted additional assumptions}. For instance, while not previously stated, an explicit assumption was documented that if the geofence is changed during the flight, the change will take place when the primary controller is the safety pilot, and that if it is not feasible, the pilot will immediately be alerted. 

There were other elements beyond interfaces that were not explicitly defined in the STAMP Model. For example, there is a data log for the flight test and a requirement that all data during the flight test be recorded for future analysis. This equates to a possible UCA for every signal within the wingman where any signal that is \textit{not provided} to the data recorder, results in hazard [H5]. This was captured at the top level rather than at each wingman level 2 signal STPA analysis. 
A summary of the findings from two signals are presented here, while the more extensive STPA is listed in the appendix on the ArXiv preprint of this paper.%\footnote{Since the AIAA SciTech conference paper is limited to 20 pages, the appendix in the ArXiv preprint summarizes the STPA work.}

\subsubsection{Signal 5: Lead to Wingman}

Note that in operations, a lead would command the wingman to a position. In practice during the flight test, there are two people on the wingman aircraft - a test pilot/engineer and a safety pilot. The test pilot on the wingman aircraft acts on behalf of the lead during flight test by selecting and commanding test points. A position report is still sent from the wingman to the lead, but elements like the commanded rejoin point and test point ID are actually set on the wingman aircraft. %As an intermediary between the operational and test scenarios, the “lead” is still considered to send test points and rejoin point commands to the wingman for the purposes of the safety evaluation. 

The requirements for the signal going from the lead to the wingman (signal 5 at level 1) tend to get fairly repetitive and fit general categories. The full list is provided in the appendix on the ArXiV preprint. There are two main signals that are sent from the Lead to the wingman: a lead position report and voice coordination / safety concerns. The position report is made up of several sub-elements like timestamps, state variables, testpoint IDs, etc. For each of these cases, the two applicable UCAs were that the signal was provided when it was inaccurate or that the signal was not provided. For each case, a standard set of requirements were generated. In cases where an inaccurate signal is provided, the following requirements are applicable (a generic [variable] is used to represent that each variable output from the lead could be inserted into that requirement  statement):
\begin{itemize}
    \item Instrument accuracy shall be reasonably checked prior to takeoff (e.g through regular inspection/calibration intervals). 
    \item The lead aircraft shall monitor sensors for errors.
    \item The lead aircraft shall perform a reasonableness check on the [variable] before transmitting. 
    \item If the lead discovers a sensor error or unreasonable value for [variable], it shall be communicated as an invalid [variable] in the position report. 
    \item The wingman aircraft shall perform a reasonableness check on the lead [variable]. 
    \item If the error is outside acceptable operational bounds, the wingman shall follow an appropriate contingency plan.
    \item In the cases where a signal was not provided, the following requirements were applicable:
    \begin{itemize}
        \item Instrument accuracy shall be reasonably checked prior to takeoff (e.g through regular inspection/calibration intervals). 
        \item The lead aircraft shall monitor sensors for errors.
        \item If the wingman uncertainty for the lead aircraft's [variable] is outside acceptable operational bounds or cannot be estimated given a history of other data, the wingman shall follow an appropriate contingency plan.
    \end{itemize}
\end{itemize}
Coming up with contingency plans for each scenario is a non-trivial task in itself that will generate additional requirements for the system. 
Additional specific requirements related to timing included:
\begin{itemize}
    \item RL1.5.3 The elements of the position report for the lead shall be sampled at the same time step. (Synchronous Communications)
    \item RL1.5.4 The wingman's operation shall have reasonable tolerance for small delays in position updates from the lead. (Robustness to latency)
\end{itemize}

Requirements on how the wingman acted in response to the commanded rejoin point included: 
\begin{itemize}
    \item RL1.5.5a The wingman shall not move to a rejoin point that violates safety.
    \item RL1.5.5b The lead shall double check that the correct rejoin point is sent (e.g. as a checklist item). Alternatively, verification could be used to indicate that the error rate of the commanded rejoin point is below some threshold.
\end{itemize}
Requirements RL1.5.36-RL1.5.42 related to voice commands included:
\begin{itemize}
    \item All flight personnel shall be briefed on communication protocol before takeoff. 
    \item Communication systems shall be tested before takeoff to ensure they work (e.g through regular inspection/calibration intervals).
    \item Measures shall be taken to fix dropouts if possible. 
    \item If the dropout can't be fixed, a contingency plan shall be followed.
    \item All flight personnel shall also be trained to recognize unsafe situations and be in contact with someone else who can (recognize unsafe situations) if they are unable to recognize unsafe situations. 
    \item If the wingman pilot needs more detailed information, they shall request it from the lead if possible. 
    \item If the lead is sharing information too early, they shall stop and notify the other parties, and then repeat the information at the appropriate time. 
    \item If the wingman tries to act on premature information, an effort shall be made to correct the mistake.
    \item If the lead fails to share information on time, the wingman shall make a best effort to maintain safe flight if necessary and a contingency plan shall be followed.
\end{itemize}

\subsubsection{L2W2: Level 2 Wingman Signal 2: Run Time Assurance to Control Selector}
The STPA analysis process highlighted specific design questions to be answered later in the program. For example, in the process of designing the RTA to respond within the EPM constraints, how hard is it to also add EPM constraints and/or SSCs to the RTA? Additionally, how much buffer is needed to reduce the number of times that the EPM switches control to the safety pilot and terminates testing of the NNCS/RTA?

There were four primary categories of requirements on signal from the RTA to the CS (W2):
\begin{enumerate}
    \item If the NNCS command would cause the wingman aircraft to [condition], the RTA shall correct the control output to prevent [condition]. 
    \begin{itemize}
        \item violate safe separation / violating safe separation
        \item exit the approved airspace / exiting the approved airspace
    \end{itemize}
    \item The RTA shall perform maneuvers that adhere to safety constraints.
    \begin{itemize}
        \item Stays within structural limits of the aircraft
        \item Stays within the aircraft flight envelope
        \item Stays within pilot physiological limits (flight test only if this is eventually intended for unmanned aircraft)
    \end{itemize}
    \item The RTA shall only activate when necessary.
    \item The RTA fails to produce an output, cannot produce a safe output, or suffers some other type of fault where it fails, a contingency plan shall be followed.
    \begin{itemize}
        \item If the RTA fails to correct the command from the NNCS due to error, the RTA shall inform the human pilot (via W12) and shall send an error signal to the CS (W2), which will automatically switch control to the human pilot (W5$\to$W3).
        \item If the RTA fails to correct the command from the NNCS due to error, a contingency plan shall be followed by the Safety Pilot.
        \item If the RTA provides a command that violates SSCs, it is up to the discretion of the human safety pilot to intervene. 
        \item The human safety pilot shall be made aware that the RTA does not protect against SSC violations.
    \end{itemize}
\end{enumerate}
Because no system is perfect, values for acceptable rates of missed detections (not correcting unsafe action) and false alarms (correcting an action that was safe) should be determined during the design process. So should required frequencies for the RTA to operate at, as well as the the robustness required of the RTA to noise and latency.

%%%%%%%%%%%%%%%%%%%%% Conclusions %%%%%%%%%%%%%%%%%%%%%%%%%%%%%%%%%%%%%%%%%%%%%
\section{Conclusion} \label{Conclusion}
RTA will be a critical tool to assuring safety of NNCS on robotic systems like aircraft. This research applied STAMP and STPA to identify safety requirements for an RTA system bounding an NNCS. The work represented the first application of STAMP and STPA to an NNCS bounded by RTA, the first functional control block diagram to contain a combination of EPM and RTA for flight testing an NNCS on a full scale aircraft, and a resulting set of accidents, hazards, safety constraints, SSCs, common cause scenarios, UCAs, scenarios and safety requirements to inform verification and validation of an NNCS bounded by RTA. In the future, this research can be expanded to operational scenarios with more wingman aircraft and a larger scope of foreseeable operating conditions. Additionally, the team is using the output of the STAMP and STPA to argue safety of an NNCS bounded by RTA through an assurance case \cite{rinehart2015current}.

\section*{Acknowledgments}
The authors would like to thank Karl Salva, Dan Montes, Kevin Price, Jason Shroyer, Tyler "Kode" Brown, Cody Flemming, David Kapp, Ray Garcia, Natasha Neogi, and others for their input and feedback on this work. This work was supported by the Air Force Research Laboratory's Sensors and Aerospace Systems Directorates, the University of Cincinnati Women in Science and Engineering Program, the Test Resource Management Center, and the Air Force Research Laboratory ADIDRUS Contract. The views expressed are those of the authors and do not reflect the official guidance or position of the United States Government, the Department of Defense or of the United States Air Force.

\bibliography{references}

%%%%%%%%%%%%%%%%%%%%% Accidents, Hazards, Safety Constraints %%%%%%%%%%%%%%%%%%%%%%%%%%%%%%%%%%%%%%%%%%%%%
%\newpage

\section*{Appendix: STPA Tables}
The unsafe control actions, scenarios, and requirements in this section represent the current work in progress over a multi-year effort. As such, requirements in particular are at an early research stage, and alternative requirements may also meed the needs of preventing unsafe control actions. 

\subsection{L2W1: Level 2 Wingman Signal 1: Neural Network to Run Time Assurance}
The NNCS receives lead and wingman state information from the airframe and computes a set of control commands that are sent to the RTA block.

\begin{table}[hbt!]
\begin{center}
\caption{\label{t:STPA_W1} W1: Neural Network Control System to Run Time Assurance}
\begin{tabular}[t]{|L{4.5cm} | L{5cm} | L{5.5cm} |}
\hline
\textbf{Unsafe Control Action}	&	\textbf{Rationale / Scenario}	&	\textbf{Requirement}	\\\hline
NNCS \textbf{provides} a command to the RTA that would a loss of safe separation [H1], loss of control [H2], violation of geofence [H3], or harm to the pilot or aircraft [H4], or violate EPM limits, terminating testing early [H6].   
    & This could occur due to training that does not consider safety constraints.
	& [RL2.W1.1] The NNCS shall be trained with penalties that account for soft safety constraints and EPM limits, and/or an RTA should be used during the training process.  \\\hline
NNCS \textbf{does not provide} a control command to the RTA.   
    & This could occur due to a fault in the NNCS component.
	& [RL2.W1.2] In the presence of an NNCS fault, a default value shall be output from the NNCS.  \\\hline
	& & [RL2.W1.3] The pilot shall be notified of NNCS faults (via W13).\\\hline 
NNCS \textbf{does not provide} data to be recorded to the on-board database [H5].
    & Failing to record NNCS data could happen due to a design flaw or running out of time within the timeframe to send the data. 
	& [RL2.W1.4]The data recorded about NNCS activation shall be identified and checked. It should also be determined whether the NNCS has sufficient time within a specific operation frame on operational hardware to complete it's function as well as send data to be recorded. If the data cannot be recorded, the pilot or test engineer shall be alerted.  \\\hline
NNCS \textbf{is too early or too late to provide} a control command to the RTA, resulting in test failure [H6] or flight instability [H2].   
    & NNCS could send an early or late signal if the NNCS falls out of sync with the airframe and RTA.  
	& [RL2.W1.5] The NNCS, Airframe, and RTA shall be synchronized (e.g. reference the same clock)\\\hline
\end{tabular}
\end{center}
\end{table}

\newpage
\subsection{L2W2: Level 2 Wingman Signal 2: Run Time Assurance to Control Selector}
%The RTA receives an observation of the current aircraft state from the airframe, on/off and settings from the operator, and control commands from the NNCS. The RTA sends a filtered pitch, roll, speed, and yaw control commands that ensure collision avoidance (safe separation) and geofence satisfaction to the CS; status to the pilot/engineer. %
%
The RTA is responsible for assuring collision freedom and geofence adherence within the limits if the EPM. %The RTA does not monitor soft safety constraints. The proposed RTA design will begin by focusing on combining geofence and RTA as an initial proof of concept; however, future RTA designs should consider inclusion of jetwash avoidance. As the design currently stands, the responsibility of jetwash avoidance is placed on the human safety pilot. The EPM/Switch might also have a requirement to check reasonableness of input from RTA in the event that there is corruption between the RTA output and input to the EPM/Switch.

\begin{table}[hbt!]
\begin{center}
\caption{\label{t:STPA_W2_provides1} W2: Unsafe control actions between the RTA and Control Selector: Provides (1 of 2)}
\begin{tabular}[t]{|L{4.5cm} | L{5cm} | L{5.5cm} |}
\hline
\textbf{Unsafe Control Action}	&	\textbf{Rationale / Scenario}	&	\textbf{Requirement}	\\\hline
RTA \textbf{provides} a control command to control selector when the command violates safe separation [H1], causes loss of control [H2], violates geofence [H3], causes harm to the pilot or aircraft [H4].
    & Sending an unsafe command could occur because the aircraft is outside of expected operational conditions. For example, the angle of attack of the aircraft could be outside the RTA design range the RTA.
	& [RL2.W2.1] The RTA shall be designed to provide geofence and collision avoidance within the entire range of EPM limits.	\\\hline
     & Sending an unsafe command could occur due to a fault in the lead position report, wingman position report, or entity reports (a fault in W4). 
	& [RL2.W2.2] The RTA shall check the reasonableness of inputs based on history and aircraft limits. If an unreasonable input is found, the RTA shall alert the safety pilot that the RTA is in a failed state via W7.  \\\hline
	& Sending an unsafe command could occur because no signal comes from the NNCS due to a fault (a fault in W1). 
	& [RL2.W2.3] In the absence of a signal from the NNCS, the RTA should use the last control command from the NNCS in the computation of an RTA response.   \\\hline
	& Sending an unsafe command could occur because no signal was ever output from the NNCS.  
	& [RL2.W2.4] When no last control command from the NNCS is available, RTA shall assume a maintain maneuver command based on the current state of the aircraft.   \\\hline
    & Sending an unsafe command could occur due to a fault the specification of the safe separation distance or geofence.
	& [RL2.W2.5] The units and data format for safe separation distance and geofence shall be reasonably obvious to the operator. Safe separation and geofence values input by the operator shall be checked for a reasonableness.\\\hline
\end{tabular}
\end{center}
\end{table}

\begin{table}[hbt!]
\begin{center}
\caption{\label{t:STPA_W2_provides2} W2: Unsafe control actions between the RTA and Control Selector: Provides (2 of 2)}
\begin{tabular}[t]{|L{4.5cm} | L{5.5cm} | L{5cm} |}
\hline
\textbf{Unsafe Control Action}	&	\textbf{Rationale / Scenario}	&	\textbf{Requirement}	\\\hline
RTA \textbf{provides} a control command to control selector when the command violates safe separation [H1], causes loss of control [H2], violates geofence [H3], causes harm to the pilot or aircraft [H4].
	& Sending an unsafe command could occur due to a fault in the RTA computation. 
	& [RL2.W2.6] The RTA computation shall be dual redundant \textit{and / or} checked for reasonableness properties \textit{and / or} the human safety pilot shall be prepared to take control during the test.\\\hline
	& Sending an unsafe command could occur due to a fault in transmission to the CS.
	& [RL2.W2.7] The output of the RTA shall be shall be dual redundant \textit{and / or} checked for reasonableness properties. 	\\\hline
	& Sending an unsafe command could occur due to a CS/RTA mismatch in units or data format. For example, a speed command given in ft/s when knots is expected. 
	& [RL2.W2.8] The output of the RTA shall adhere to the expected units and format of the CS.  		\\\hline
	& If the output trajectory puts the aircraft in the jetwash of the lead aircraft for some minimum amount of time, it could case the aircraft to lose control. 
	& [RL2.W2.9] Knowledge of jetwash is outside the scope of the RTA, and trajectories towards the jetwash of the lead aircraft should be monitored for safety by the safety pilot.	\\\hline
RTA \textbf{provides} a control command to control selector when the command violates the envelope monitored by the EPM, resulting in a loss of planned test [H6], or potential harm to the human pilot or aircraft [H4].
    & Sending an command outside the EPM unsafe command could occur because the aircraft is outside of expected operational conditions. For example, the angle of attack of the aircraft could be outside the range the RTA is designed for.
	& [RL2.W2.10] The RTA shall be designed to provide geofence and collision avoidance within the EPM limits.	\\\hline
\end{tabular}
\end{center}
\end{table}

\begin{table}[hbt!]
\begin{center}
\caption{\label{t:STPA_W2_too_late} W2: Unsafe control actions between the RTA and Control Selector: Timing (too late)}
\begin{tabular}[t]{|L{4.5cm} | L{5.5cm} | L{5cm} |}
\hline
\textbf{Unsafe Control Action}	&	\textbf{Rationale / Scenario}	&	\textbf{Requirement}	\\\hline
RTA \textbf{provides a late} a control command to control selector when necessary to provide safe separation [H1] or stay in geofence [H3].
    & RTA could send a late signal if the RTA and CS fall out of sync.  
	& [RL2.W2.11] The RTA and CS shall be synchronized (e.g. reference the same clock)\\\hline
RTA \textbf{provides a late} a control command to control selector when necessary to provide safe separation [H1] or stay in geofence [H3].
    & RTA could send a late signal if the computation time of the solution causes a frame overrun.
	& [RL2.W2.12] The RTA shall be able to compute a solution within a portion of the allotted frame rate.\\\hline
\end{tabular}
\end{center}
\end{table}

\newpage
\begin{table}[hbt!]
\begin{center}
\caption{\label{t:STPA_W2_does_not_provide} W2 Unsafe control actions between the RTA and Control Selector: Does Not Provide}
\begin{tabular}[t]{|L{4.5cm} | L{5.5cm} | L{5cm} |}
\hline
\textbf{Unsafe Control Action}	&	\textbf{Rationale / Scenario}	&	\textbf{Requirement}	\\\hline
RTA \textbf{does not provide} a control command to control selector when necessary to provide safe separation [H1] or stay in geofence [H3].
    & Failing to send control commands to assure collision protection and geofence could happen because of a data dropout between the RTA and the CS. 
	&  [RL2.W2.13] The CS shall alert the pilot (via W14) when no signal is received (from W2).\\\hline
    & Failing to send control commands to assure collision protection and geofence could happen because it is not possible to send a safe signal. 
	& [RL2.W2.14] The RTA should alert the pilot of a failure of the RTA is a safe command cannot be found.\\\hline
	& This could occur if the pilot or test engineer turns off the RTA and then turns it back on again mid-flight when it is already violating safe separation or geofence. 
	&  [RL2.W2.15] The interface to the pilot/test engineer will check if the geofence or safe separation is being violated when RTA is off, and will alert the pilot/test engineer before they turn the RTA back on. \\\hline
RTA \textbf{does not provide} data to be recorded to the on-board database [H5].
    & Failing to record RTA data could happen due to a design flaw or running out of time within the timeframe to send the data. 
	& [RL2.W2.16] The data recorded about RTA activation shall be identified and checked. It should also be determined whether the RTA has sufficient time within a specific operation frame on operational hardware to complete it's function as well as send data to be recorded. If the data cannot be recorded, the pilot or test engineer shall be alerted.  \\\hline
\end{tabular}
\end{center}
\end{table}

\newpage
\subsection{L2W3: Level 2 Wingman Signal 3: Control Switch to Airframe}
The control switch selects whether to send the NNCS/RTA or safety pilot signal to the airframe.

\begin{table}[hbt!]
\begin{center}
\caption{\label{t:STPA_W3} W3: Unsafe control actions between the CS and Airframe}
\begin{tabular}[t]{|L{4.5cm} | L{5cm} | L{5.5cm} |}
\hline
\textbf{Unsafe Control Action}	&	\textbf{Rationale / Scenario}	&	\textbf{Requirement}	\\\hline
CS \textbf{provides} a control command to the airframe when the command causes loss of control [H2] or causes harm to the pilot or aircraft [H4].
	& Sending an unsafe command could occur due to a fault in the CS switching. 
	& [RL2.W3.1] The CS should default to safety pilot control in the event of a fault.\\\hline
	& Sending an unsafe command could occur due to a fault in transmission to the airframe.
	& [RL2.W3.2] The output of the CS shall be shall be at least dual redundant \textit{and / or} checked for reasonableness properties. 	\\\hline
	& Sending an unsafe command could occur due to a CS/airframe mismatch in units or data format. For example, a speed command given in ft/s when knots is expected. 
	& [RL2.W3.3] The output of the CS shall adhere to the expected units and format of the airframe.  		\\\hline
	& Sending an unsafe command could occur because the aircraft is outside of expected operational conditions. For example, the angle of attack or roll angle could be outside the safe set for flight test.
	& [RL2.W3.4] The safety pilot shall provide control signals the keep the aircraft within safe operating conditions.	\\\hline
	& Sending an unsafe command could occur because an unsafe command came from the RTA due to a fault (a fault in W2). 
	& [RL2.W3.5] The EPM and Safety pilot shall provide sufficient redundant safety monitoring to switch when needed to mitigate RTA signal risks.   \\\hline
	& Sending an unsafe command could occur because no signal was ever output from the RTA.  
	& [RL2.W3.6] In the absence of a signal from the RTA, the CS shall default switch to safety pilot control.    \\\hline
CS \textbf{provides a late} a control command to airframe, resulting in a loss of control [H4].
    & CS could send a late signal if the CS and airframe fall out of sync.  
	& [RL2.W3.7] The CS and airframe shall be synchronized (e.g. reference the same clock)\\\hline
CS \textbf{does not provide} a control command to airframe resulting in a loss of control.
    & Failing to send control commands could happen because of a data dropout between the CS and the airframe. 
	& [RL2.W3.8] The signal from CS to the airframe shall be as reliable as practical through means such as redundancy.\\\hline
CS \textbf{does not provide} data to be recorded to the on-board database [H5].
    & Failing to record CS data could happen due to a design flaw or running out of time within the timeframe to send the data. 
	& [RL2.W3.9] The data recorded about CS state activation shall be identified and checked. It should also be determined whether the CS has sufficient time within a specific operation frame on operational hardware to complete it's function as well as send data to be recorded. If the data cannot be recorded, the pilot or test engineer shall be alerted.  \\\hline
\end{tabular}
\end{center}
\end{table}

\newpage
\subsection{L2W4: Level 2 Wingman Signal 4: Airframe State Output}
The Airframe outputs an updated state observation to the NNCS, RTA, EPM, Safety Pilot, and Test Engineer. Note that a redundant, heterogeneous source of truth is available to the the Safety Pilot and Test Engineer, because it is assumed that they have insight, beyond what is displayed in state information sensed by the airframe, i.e. the pilots can use what they see out the window as a sanity check to instrument readings.

\begin{table}[hbt!]
\begin{center}
\caption{\label{t:STPA_W4} W4: Airframe output}
\begin{tabular}[t]{|L{4.5cm} | L{5cm} | L{5.5cm} |}
\hline
\textbf{Unsafe Control Action}	&	\textbf{Rationale / Scenario}	&	\textbf{Requirement}	\\\hline
Airframe \textbf{provides} an incorrect estimation of state, which when acted on by the NNCS/RTA resulting in a control command that violates safe separation [H1], causes loss of control [H2], violates geofence [H3], causes harm to the pilot or aircraft [H4].
    & This could happen due to an error in an airframe sensor. 
	& [RL2.W4.1] The safety pilot shall check instrument readings and use visual situation assessment to provide redundant collision avoidance, geofence, and other safety monitoring.\\\hline
Airframe \textbf{does not provide} a state estimate through W4.
	& This could happen due to a fault or frame overrun in the estimation software, or a faulty sensor.
	& [RL2.W4.2] The NNCS and RTA shall be designed to experience a reasonable amount of noise or dropouts in the sensed state, also see previous requirement.\\\hline
Airframe \textbf{too late to provide} a state estimation through W4.   
    & The Airframe could send a signal too late if it falls out of with the NNCS, RTA, EPM or CS.  
	& [RL2.W4.3] The NNCS, Airframe, RTA, CS, and EPM shall be synchronized (e.g. reference the same clock)\\\hline
\end{tabular}
\end{center}
\end{table}

\newpage
\subsection{L2W5: Level 2 Wingman Signal 5: Pilot to CS}
The pilot is expected to constantly provide a backup signal to the CS parallel to the NNCS/RTA path, as well as to control the aircraft between test points.

\begin{table}[hbt!]
\begin{center}
\caption{\label{t:STPA_W5} W5: Pilot to Control Selector}
\begin{tabular}[t]{|L{4.5cm} | L{5cm} | L{5.5cm} |}
\hline
\textbf{Unsafe Control Action}	&	\textbf{Rationale / Scenario}	&	\textbf{Requirement}	\\\hline
Pilot \textbf{provides} a control command to CS that violates safe separation [H1], causes loss of control [H2], violates geofence [H3], causes harm to the pilot or aircraft [H4].
    & Sending an unsafe command could occur because the pilot is incapacitated or ability to react is impaired during the test (for example, the primary controller technically stays within safe boundaries but still imparts physiological stress). 
	& [RL2.W5.1] The safety pilot shall evaluate their health status prior to flight and terminate the NNCS/RTA control early if they suspect reasonable chance of impairment.\\\hline
	Pilot \textbf{provides} a control command to CS that violates safe separation [H1].
	& Sending an unsafe command could occur because the pilot has lost situational awareness (line of sight) to the lead.
	& [RL2.W5.2] The safety pilot shall maintain line of sight or intervene when line of sight is lost long enough for a potential collision [SSC1, SSC2].\\\hline
Pilot \textbf{does not provide} a control command to CS when needed to prevent loss of safe separation [H1], loss of control [H2], violation of geofence [H3], or harm to the pilot or aircraft [H4].
    &  This could occur due to a system fault that prevents transmission of the pilot command to the CS. 
	&  [RL2.W5.3] The signal from the pilot to the CS shall be as reliable as possible, likely having two to four redundant paths.\\\hline
Pilot \textbf{is too late to provide} a control command to prevent loss of safe separation [H1], loss of control [H2], violation of geofence [H3], or harm to the pilot or aircraft [H4].
    &  This could happen if the pilot safety pilot loses situation awareness due to distraction or loss of line of sight [SSC1,SSC2].
	&  [RL2.W5.4] The pilot should terminate the test if they become distracted or lose line of sight for too long.\\\hline
Pilot \textbf{is too early to provide} a control command to CS, when the NNCS/RTA is safe causing premature test termination and a loss of planned operations [H6].
    &  This could occur if the pilot is not well versed in the RTA and EPM safety limits prior to the flight.
	&  [RL2.W5.5] The RTA and EPM safety limits shall be included in the pre-brief for every test.\\\hline
\end{tabular}
\end{center}
\end{table}

\newpage
\subsection{L2W6: Level 2 Wingman Signal 6:Envelope Protection Monitor to Control Selector}
The EPM provides a signal to the CS to switch from NNCS/RTA input to Safety Pilot input when commanded by the Safety pilot (via W9) or EPM limits are violated.
\begin{table}[hbt!]
\begin{center}
\caption{\label{t:STPA_W6} W6: Envelope Protection Monitor to Control Selector}
\begin{tabular}[t]{|L{4.5cm} | L{5cm} | L{5.5cm} |}
\hline
\textbf{Unsafe Control Action}	&	\textbf{Rationale / Scenario}	&	\textbf{Requirement}	\\\hline
EPM \textbf{provides} a switch command to the CS when the pilot has not commanded it and no safety criteria are violated, resulting a a loss of test point data generation [H6].
    & This could occur when the aircraft is close to safety violations and there is noise. 
	& [RL2.W6.1] The NNCS should be trained to stay within EPM limits under noise and the RTA stay within EPM limits under noise.\\\hline
EPM \textbf{does not provide} a switch command to the CS when the pilot has commanded it or safety criteria are violated, resulting a loss of safe separation [H1], loss of control [H2], violation of geofence [H3], or harm to the pilot or aircraft [H4].
    & This could occur due to a fault in the EPM. 
	& [RL2.W6.2] The pilot shall be informed of any faults in the EPM (W8), and the EPM switch should default to output a switch to pilot control in the presence of faults.\\\hline
EPM \textbf{is too early to command} a switch command to the CS to pilot control when no safety criteria are violated, resulting a a loss of test point data generation [H6].
    & This could occur when the aircraft is close to safety violations and there is noise. 
	& [RL2.W6.3] The NNCS should be trained to stay within EPM limits under noise and the RTA stay within EPM limits under noise.\\\hline
EPM \textbf{is too late to command} a switch command to the CS when the pilot has commanded it or safety criteria are violated, resulting a loss of safe separation [H1], loss of control [H2], violation of geofence [H3], or harm to the pilot or aircraft [H4].
    & This could occur due to a fault in the EPM. 
	& [RL2.W6.4] The pilot shall be informed of any faults in the EPM (W8), and the EPM switch should default to output a switch to pilot control in the presence of faults.\\\hline
\end{tabular}
\end{center}
\end{table}

\newpage
\subsection{L2W8: Level 2 Wingman Signal 8}
\begin{table}[hbt!]
\begin{center}
\caption{\label{t:STPA_W8} W8: Envelope Protection Management to Pilot}
\begin{tabular}[t]{|L{4.5cm} | L{5cm} | L{5.5cm} |}
\hline
\textbf{Unsafe Control Action}	&	\textbf{Rationale / Scenario}	&	\textbf{Requirement}	\\\hline
EPM provides engaged/disengaged status to Pilot when the status is incorrect.	&	An error causes the EPM to provide the wrong status. If the EPM reports the status as engaged when it is not, the pilot may try to take control of the aircraft when they should not, which could jeopardize the flight test and the process of collecting test data. It may also alarm the pilot if it happens unexpectedly or if they are unable to take control, causing them to call off the flight entirely. If the EPM reports the status as disengaged when it is engaged, the pilot may not know they need to control the aircraft and it could start to lose control.	
& [RL2.W8.1] The EPM shall report the correct status to the pilot. If error caused the EPM to report the wrong status, an effort shall be made to correct the error. If it cannot be corrected, a contingency plan shall be followed	\\\hline
EPM does not provide engaged/disengaged status to Pilot when the pilot needs to know the status.	
&	An error causes the EPM to not provide the status or an error causes the status to not be displayed, even if. If the pilot does not know the status when the EPM is engaged, the pilot may not know they need to control the aircraft and the aircraft could start to lose control. If the pilot does not know the status when the EPM in disengaged, they may not be sure if the aircraft is being controlled and they may feel unsafe, which might cause them to call off the flight.	
& [RL2.W8.2] The EPM shall report its status to the pilot. If error causes it to not provide its status, an effort shall be made to correct the error. If the error cannot be corrected, a contingency plan shall be followed.	\\\hline
\end{tabular}
\end{center}
\end{table}
\subsection{L2W14: Level 2 Wingman Signal 14}
 The CS shall alert the pilot (via W14) when no signal is received (from W2).
 
\newpage
 \subsection{L1.5: Level 1 Signal 5: Lead to Wingman}
The EPM provides a signal to the CS to switch from NNCS/RTA input to Safety Pilot input when commanded by the Safety pilot (via W9) or EPM limits are violated.
\begin{table}[hbt!]
\begin{center}
\caption{\label{t:STPA_:L151} Level 1 Signal 5: Lead to Wingman (1)}
\begin{tabular}[t]{|L{3.5cm} | L{6cm} | L{5.5cm} |}
\hline
\textbf{Unsafe Control Action}	&	\textbf{Rationale / Scenario}	&	\textbf{Requirement}	\\\hline
Lead provides its position report to Wingman when it is inaccurate.	&	Data in position report could be corrupt or incorrect due to sensor error. If the wingman doesn't have an accurate position for the lead, it may fail to maintain safe separation [CCS1, CCS2], or it may try to get to a rejoin position that corresponds with the inaccurate report [H6/SC6], or it may fly through the jet wash [CCS5] and lose control.	
& [RL1.5.1]	There shall be a reasonableness check on the wingman and/or on the lead that the position report from the lead is feasible. If the wingman receives an invalid report, it shall say why and shall notify the proper controller (lead or ground station) and proceed with the proper contingency plan (maintain current flight path, return to base, loiter, etc.)	\\\hline

Lead does not provide its position report to Wingman when the wingman is in close proximity to the lead.	&	The position report might be missing because of a system error or data dropout. If the wingman doesn't have the position of the lead, it may fail to maintain safe separation [CCS1, CCS2], or it may fly through the jet wash [CCS5] and lose control.	
&  [RL1.5.2]	The wingman shall alert the flight lead pilot / and operator that it does not know where the lead is and proceed with the proper contingency plan (maintain current flight path, return to base, loiter, fly to a distant point, etc.)	\\\hline

Lead provides its position report to Wingman too early or too late when not all elements (ex. position, velocity, attitude) are up to date.	&	There could be a fault in a sensor on the lead, that is providing asynchronous updates to elements of the position report (e.g updating position and velocity at different rates), where one update doesn't make it in the frame.	& [RL1.5.3]	The elements of the position report for the lead shall be sampled at the same time step.	\\\hline

Lead provides its position report to Wingman too late when the wingman is on a collision course.	&	There could be a fault in a sensor on the lead, that is providing asynchronous updates to elements of the position report (e.g updating position and velocity at different rates), where one update doesn't make it in the frame.	& [RL1.5.4]	The wingman's operation shall have reasonable tolerance for small delays in position updates from the lead.	\\\hline

Lead (test pilot on wingman aircraft during test) provides a commanded rejoin point to Wingman when it is incorrect.	&	The data coming from the lead could be corrupted or the lead could have accidentally commanded an incorrect point.	& [RL1.5.5]	"The wingman shall not move to a rejoin point that violates safety.
The lead shall double check that the correct rejoin point is sent (e.g. as a checklist item)."	\\\hline

Lead (test pilot on wingman aircraft during test) does not provide a commanded rejoin point to Wingman when a coordinated test point has started.	&	It's possible the lead never sent the signal or that there is a data dropout.	& [RL1.5.6]	The wingman shall communicate that they do not know the rejoin point and follow the appropriate contingency plan.	\\\hline

\end{tabular}
\end{center}
\end{table}

\begin{table}[hbt!]
\begin{center}
\caption{\label{t:STPA_:L152} Level 1 Signal 5: Lead to Wingman (2)}
\begin{tabular}[t]{|L{4.5cm} | L{5cm} | L{5.5cm} |}
\hline
\textbf{Unsafe Control Action}	&	\textbf{Rationale / Scenario}	&	\textbf{Requirement}	\\\hline
Lead provides the state timestamp to Wingman when it is incorrect.	&	There could be an error in compiling the report in which the time is incorrect or isn't updated for the system. One of the aircraft may have taken off with the incorrect time.	& [RL1.5.7]	The system shall be set with the correct time before takeoff and have a method of ensuring synchronization of time during the flight.	\\\hline

Lead does not provide the state timestamp to Wingman when the wingman is in close proximity to the lead.	&	The timestamp could be left out because of a system error that leads to it not being reported or that prevents the time from being read (i.e. clock stops working). If the wingman receives a correct report without a timestamp, it won't know if that info is accurate or not, so it won't know where the lead is. Also may not be able to use the report at all, even if the rest is correct.	&[RL1.5.8]	"Instrument accuracy shall be reasonably checked prior to takeoff (e.g through regular inspection/calibration intervals). 
The lead aircraft shall check that the position report has all observations before being sent. 
If observations are missing, controllers shall be notified which are missing and why (if possible) and the lead shall attempt to send the report again. 
If errors continue to prevent attempts, steps shall be taken to correct the errors or follow an appropriate contingency plan."	\\\hline

Lead (test pilot on wingman aircraft during test) provides the test point id to Wingman when it is incorrect.	&	There was an error in the test plan or in the lead setting the test point, so the lead believes the wingman will follow one test point, while the wingman receives instructions to follow another.	&[RL1.5.9]	"Both the lead and the wingman shall have a test plan with the correct test point ids before takeoff. 
If the current id is wrong because of an error or a mistake, measures shall be taken to correct it and the report shall be sent again. If it can't be corrected, a contingency plan shall be followed."	\\\hline

Lead (test pilot on wingman aircraft during test) does not provide the test point id to Wingman when the wingman needs the test point.	&	The test point doesn't get sent because of an error or oversight. The wingman may not be able to confirm that the lead is on the same test point and may not be able to complete the test point. Or the mixup may cause delays that reduce the total number of points that can be tested in a flight.	& [RL1.5.10]	"The lead shall have the test point ids before takeoff. 
The lead aircraft shall send test points to the wingman as planned."	\\\hline

\end{tabular}
\end{center}
\end{table}

\begin{table}[hbt!]
\begin{center}
\caption{\label{t:STPA_:L153} Level 1 Signal 5: Lead to Wingman (3)}
\begin{tabular}[t]{|L{4.5cm} | L{5cm} | L{5.5cm} |}
\hline
\textbf{Unsafe Control Action}	&	\textbf{Rationale / Scenario}	&	\textbf{Requirement}	\\\hline
Lead provides its position to Wingman when it is incorrect.	&	The data could be corrupt due to transmission error, sensor error, or excess noise.	& [RL1.5.11]	"Instrument accuracy shall be reasonably checked prior to takeoff (e.g through regular inspection/calibration intervals). 
The lead aircraft shall monitor sensors for errors.
The lead aircraft shall perform a reasonableness check on the position before transmitting.
If the lead discovers a sensor error or unreasonable value for position, it shall be communicated as an invalid position in the position report.
The wingman aircraft shall perform a reasonableness check on the lead position. 
If the error is outside acceptable operational bounds, the wingman shall follow an appropriate contingency plan."	\\\hline

Lead does not provide its position to Wingman when the wingman is in close proximity to the lead.	&	The position report might be missing because of a system error or data dropout. If the wingman doesn't have the position of the lead, it may fail to maintain safe separation [CCS1, CCS2], or it may fly through the jet wash [CCS5] and lose control.	&	[RL1.5.12]	The wingman shall alert the flight lead pilot / and operator that it does not know where the lead is and proceed with the proper contingency plan.	\\\hline

Lead provides its orientation to Wingman when it is incorrect.	&	The orientation could be incorrect because of excess sensor noise, sensor error, or an error that changed the data in transmission.	& [RL1.5.13]	"Instrument accuracy shall be reasonably checked prior to takeoff (e.g through regular inspection/calibration intervals). 
The lead aircraft shall monitor sensors for errors.
The lead aircraft shall perform a reasonableness check on the orientation before transmitting.
If the lead discovers a sensor error or unreasonable value for orientation, it shall be communicated as an invalid orientation in the position report.
The wingman aircraft shall perform a reasonableness check on the lead orientation. 
If the error is outside acceptable operational bounds, the wingman shall follow an appropriate contingency plan."	\\\hline

\end{tabular}
\end{center}
\end{table}

\begin{table}[hbt!]
\begin{center}
\caption{\label{t:STPA_:L154} Level 1 Signal 5: Lead to Wingman (4)}
\begin{tabular}[t]{|L{4.5cm} | L{5cm} | L{5.5cm} |}
\hline
\textbf{Unsafe Control Action}	&	\textbf{Rationale / Scenario}	&	\textbf{Requirement}	\\\hline
Lead does not provide its orientation to Wingman when the wingman is in close proximity to the lead.	&	The orientation might not have been provided due to an error preventing the sensor from collecting data or due to an error that prevented the data from being recorded and sent.	& [RL1.5.14]	"Instrument accuracy shall be reasonably checked prior to takeoff (e.g through regular inspection/calibration intervals). 
The lead aircraft shall monitor sensors for errors.
If the wingman uncertainty for the lead aircraft orientation is outside acceptable operational bounds or cannot be estimated given a history of other data, the wingman shall follow an appropriate contingency plan."	\\\hline

Lead provides its orientation rates to Wingman when they are incorrect.	&	The orientation rate could be incorrect because of sensor error or an error that changed the data. If the wingman doesn't have the correct orientation rate, it will not be able to accurately project the path of the lead into the future, and it will be more difficult for it to maneuver safely around the lead	&  [RL1.5.15]		"Instrument accuracy shall be reasonably checked prior to takeoff (e.g through regular inspection/calibration intervals). 
The lead aircraft shall monitor sensors for errors.
The lead aircraft shall perform a reasonableness check on the orientation rates before transmitting.
If the lead discovers a sensor error or unreasonable value for orientation rates, it shall be communicated as an invalid orientation rate in the position report.
The wingman aircraft shall perform a reasonableness check on the lead orientation rates. 

If the error is outside acceptable operational bounds, the wingman shall follow an appropriate contingency plan."	\\\hline
Lead provides its true air speeds to Wingman when they are incorrect.	&	The true air speeds might be incorrect because of sensor error, calculation error, or recording error. If the wingman doesn't have the correct air speeds for the lead, it won't be able to accurately keep track of the lead's position and navigation and won't be able to safely maneuver around it.	&  [RL1.5.16]		"Instrument accuracy shall be reasonably checked prior to takeoff (e.g through regular inspection/calibration intervals). 
The lead aircraft shall monitor sensors for errors.
The lead aircraft shall perform a reasonableness check on the true air speed before transmitting.
If the lead discovers a sensor error or unreasonable value for true air speed, it shall be communicated as an invalid orientation rate in the position report.
The wingman aircraft shall perform a reasonableness check on the lead true air speed. 
If the error is outside acceptable operational bounds, the wingman shall follow an appropriate contingency plan."	\\\hline

\end{tabular}
\end{center}
\end{table}

\begin{table}[hbt!]
\begin{center}
\caption{\label{t:STPA_:L155} Level 1 Signal 5: Lead to Wingman (5)}
\begin{tabular}[t]{|L{4.5cm} | L{5cm} | L{5.5cm} |}
\hline
\textbf{Unsafe Control Action}	&	\textbf{Rationale / Scenario}	&	\textbf{Requirement}	\\\hline

Lead does not provide its true air speeds to Wingman when the wingman is in close proximity to the lead.	&	The true air speeds might be incorrect because of sensor error, calculation error, or recording error. If the wingman doesn't have the correct air speeds for the lead, it won't be able to accurately keep track of the lead's position and navigation and won't be able to safely maneuver around it.	&  [RL1.5.17]		"Instrument accuracy shall be reasonably checked prior to takeoff (e.g through regular inspection/calibration intervals). The lead aircraft shall monitor sensors for errors.
If the wingman uncertainty for the lead aircraft's true airspeed is outside acceptable operational bounds or cannot be estimated given a history of other data, the wingman shall follow an appropriate contingency plan."	\\\hline

Lead provides its velocities to Wingman when they are incorrect.	&	The velocities might be incorrect because of sensor error, calculation error, or recording error. If the wingman doesn't have the correct velocities for the lead, it won't be able to accurately keep track of the lead's position and navigation and won't be able to safely maneuver around it.	&  [RL1.5.18]		"Instrument accuracy shall be reasonably checked prior to takeoff (e.g through regular inspection/calibration intervals). 
The lead aircraft shall monitor sensors for errors.
The lead aircraft shall perform a reasonableness check on the velocities before transmitting.
If the lead discovers a sensor error or unreasonable value for velocities, it shall be communicated as an invalid orientation rate in the position report.
The wingman aircraft shall perform a reasonableness check on the lead velocities. 
If the error is outside acceptable operational bounds, the wingman shall follow an appropriate contingency plan."	\\\hline

Lead does not provide its velocities to Wingman when the wingman is in close proximity to the lead.	&	The velocities might not be provided because of sensor error, calculation error, or recording error. If the wingman doesn't have the correct velocities for the lead, it won't be able to accurately keep track of the lead's position and navigation and won't be able to safely maneuver around it.	&  [RL1.5.19]		"Instrument accuracy shall be reasonably checked prior to takeoff (e.g through regular inspection/calibration intervals). 
The lead aircraft shall monitor sensors for errors.
If the wingman uncertainty for the lead aircraft's velocities is outside acceptable operational bounds or cannot be estimated given a history of other data, the wingman shall follow an appropriate contingency plan."	\\\hline

\end{tabular}
\end{center}
\end{table}

\begin{table}[hbt!]
\begin{center}
\caption{\label{t:STPA_:L156} Level 1 Signal 5: Lead to Wingman (6)}
\begin{tabular}[t]{|L{4.5cm} | L{5cm} | L{5.5cm} |}
\hline
\textbf{Unsafe Control Action}	&	\textbf{Rationale / Scenario}	&	\textbf{Requirement}	\\\hline

Lead provides its accelerations to Wingman when it is incorrect.	&	The accelerations might be incorrect because of sensor error, calculation error, or recording error. If the wingman doesn't have the correct accelerations for the lead, it won't be able to accurately keep track of the lead's position and navigation and won't be able to safely maneuver around it.	&  [RL1.5.20]		"Instrument accuracy shall be reasonably checked prior to takeoff (e.g through regular inspection/calibration intervals). 
The lead aircraft shall monitor sensors for errors.
The lead aircraft shall perform a reasonableness check on the accelerations before transmitting.
If the lead discovers a sensor error or unreasonable value for accelerations, it shall be communicated as an invalid orientation rate in the position report.
The wingman aircraft shall perform a reasonableness check on the lead accelerations. 
If the error is outside acceptable operational bounds, the wingman shall follow an appropriate contingency plan."	\\\hline

Lead does not provide its accelerations to Wingman when the wingman is in close proximity to the lead.	&	The accelerations might not be provided because of sensor error, calculation error, or recording error. If the wingman doesn't have the correct accelerations for the lead, it won't be able to accurately keep track of the lead's position and navigation and won't be able to safely maneuver around it.	&  [RL1.5.21]		"Instrument accuracy shall be reasonably checked prior to takeoff (e.g through regular inspection/calibration intervals). 
The lead aircraft shall monitor sensors for errors.
If the wingman uncertainty for the lead aircraft's accelerations is outside acceptable operational bounds or cannot be estimated given a history of other data, the wingman shall follow an appropriate contingency plan."	\\\hline

Lead provides the amount of fuel remaining to Wingman when it is incorrect.	&	There could be an incorrect reading of the weight of the fuel due to a sensor fault or an error in compiling the report. This value being incorrect is unsafe because it provides valuable information about the state of the lead and tells the wingman whether or not the lead's behavior will change. If the actual fuel level was low, the lead might need to cut the flight short or might have performance/control issues and the fuel level will provide the wingman with some warning, in addition to other methods of communication.	&  [RL1.5.22]		"Instrument accuracy shall be reasonably checked prior to takeoff (e.g through regular inspection/calibration intervals). 
The lead aircraft shall monitor sensors for errors.
The lead aircraft shall perform a reasonableness check on the amount of fuel remaining before transmitting.
If the lead discovers a sensor error or unreasonable value for amount of fuel remaining, it shall be communicated as an invalid orientation rate in the position report.
The wingman aircraft shall perform a reasonableness check on the lead amount of fuel remaining. 
If the error is outside acceptable operational bounds, the wingman shall follow an appropriate contingency plan."	\\\hline

\end{tabular}
\end{center}
\end{table}

\begin{table}[hbt!]
\begin{center}
\caption{\label{t:STPA_:L157} Level 1 Signal 5: Lead to Wingman (7)}
\begin{tabular}[t]{|L{4.5cm} | L{5cm} | L{5.5cm} |}
\hline
\textbf{Unsafe Control Action}	&	\textbf{Rationale / Scenario}	&	\textbf{Requirement}	\\\hline

Lead does not provide the amount of fuel remaining to Wingman when the wingman is in close proximity to the lead.	&	There could be a error in the sensors or in recording that causes the amount of fuel to not be captured and sent in the report.	&  [RL1.5.23]		"Instrument accuracy shall be reasonably checked prior to takeoff (e.g through regular inspection/calibration intervals). 
The lead aircraft shall monitor sensors for errors.
If the wingman uncertainty for the lead aircraft's amount of fuel remaining is outside acceptable operational bounds or cannot be estimated given a history of other data, the wingman shall follow an appropriate contingency plan."	\\\hline

Lead provides its calibrated air speed to Wingman when it is incorrect.	&	The calibrated air speed might be incorrect because of sensor error, calculation error, or recording error. If the wingman doesn't have the correct air speed for the lead, it won't be able to accurately keep track of the lead's position and navigation and won't be able to safely maneuver around it.	&[RL1.5.24]	"Instrument accuracy shall be reasonably checked prior to takeoff (e.g through regular inspection/calibration intervals). 
The lead aircraft shall monitor sensors for errors.
The lead aircraft shall perform a reasonableness check on the calibrated air speed before transmitting.
If the lead discovers a sensor error or unreasonable value for calibrated air speed, it shall be communicated as an invalid orientation rate in the position report.
The wingman aircraft shall perform a reasonableness check on the lead calibrated air speed. 
If the error is outside acceptable operational bounds, the wingman shall follow an appropriate contingency plan."	\\\hline

Lead does not provide its calibrated air speed to Wingman when the wingman is in close proximity with the lead.	& The calibrated air speed might not be provided because of sensor error, calculation error, or recording error. If the wingman doesn't have the correct air speed for the lead, it won't be able to accurately keep track of the lead's position and navigation and won't be able to safely maneuver around it.	&[RL1.5.25]	"Instrument accuracy shall be reasonably checked prior to takeoff (e.g through regular inspection/calibration intervals). 
The lead aircraft shall monitor sensors for errors.
If the wingman uncertainty for the lead aircraft's calibrated air speed is outside acceptable operational bounds or cannot be estimated given a history of other data, the wingman shall follow an appropriate contingency plan."	\\\hline

\end{tabular}
\end{center}
\end{table}

\begin{table}[hbt!]
\begin{center}
\caption{\label{t:STPA_:L158} Level 1 Signal 5: Lead to Wingman (8)}
\begin{tabular}[t]{|L{4.5cm} | L{5cm} | L{5.5cm} |}
\hline
\textbf{Unsafe Control Action}	&	\textbf{Rationale / Scenario}	&	\textbf{Requirement}	\\\hline
Lead provides its normal accelerations to Wingman when they are incorrect.	&	The normal accelerations might be incorrect because of sensor error, calculation error, or recording error. If the wingman doesn't have the correct accelerations for the lead, it won't be able to accurately keep track of and predict the lead's position and navigation and won't be able to safely maneuver around it.	& [RL1.5.26]	"Instrument accuracy shall be reasonably checked prior to takeoff (e.g through regular inspection/calibration intervals). 
The lead aircraft shall monitor sensors for errors.
The lead aircraft shall perform a reasonableness check on the normal accelerations before transmitting.
If the lead discovers a sensor error or unreasonable value for normal accelerations, it shall be communicated as an invalid orientation rate in the position report.
The wingman aircraft shall perform a reasonableness check on the lead normal accelerations. 
If the error is outside acceptable operational bounds, the wingman shall follow an appropriate contingency plan."	\\\hline

Lead does not provide its normal accelerations to Wingman when the wingman is in close proximity to the lead.	&	The normal accelerations might not be provided because of sensor error, calculation error, or recording error. If the wingman doesn't have the correct acceleration for the lead, it won't be able to accurately keep track of and predict the lead's position and navigation and won't be able to safely maneuver around it.	& [RL1.5.27]	"Instrument accuracy shall be reasonably checked prior to takeoff (e.g through regular inspection/calibration intervals). 
The lead aircraft shall monitor sensors for errors.
If the wingman uncertainty for the lead aircraft's normal accelerations is outside acceptable operational bounds or cannot be estimated given a history of other data, the wingman shall follow an appropriate contingency plan."	\\\hline

Lead provides its PLA to Wingman when it is incorrect.	&	An error could cause the PLA value to be recorded incorrectly.	& [RL1.5.28]	"Instrument accuracy shall be reasonably checked prior to takeoff (e.g through regular inspection/calibration intervals). 
The lead aircraft shall monitor sensors for errors.
The lead aircraft shall perform a reasonableness check on the PLA before transmitting.
If the lead discovers a sensor error or unreasonable value for PLA, it shall be communicated as an invalid orientation rate in the position report.
The wingman aircraft shall perform a reasonableness check on the lead PLA. 
If the error is outside acceptable operational bounds, the wingman shall follow an appropriate contingency plan."	\\\hline

\end{tabular}
\end{center}
\end{table}

\begin{table}[hbt!]
\begin{center}
\caption{\label{t:STPA_:L159} Level 1 Signal 5: Lead to Wingman (9)}
\begin{tabular}[t]{|L{4.5cm} | L{5cm} | L{5.5cm} |}
\hline
\textbf{Unsafe Control Action}	&	\textbf{Rationale / Scenario}	&	\textbf{Requirement}	\\\hline
Lead does not provide its PLA to Wingman when the lead sends a position report.	&	An error could cause the PLA value to not be recorded.	&[RL1.5.29]	"Instrument accuracy shall be reasonably checked prior to takeoff (e.g through regular inspection/calibration intervals). The lead aircraft shall monitor sensors for errors.
If the wingman uncertainty for the lead aircraft's PLA is outside acceptable operational bounds or cannot be estimated given a history of other data, the wingman shall follow an appropriate contingency plan."	\\\hline

Lead provides its orientation angles to Wingman when they are incorrect.	&	The orientation angles might be incorrect because of sensor error, calculation error, or recording error. If the wingman doesn't have the correct orientation angles for the lead, it won't be able to accurately keep track of the lead's position and navigation and won't be able to safely maneuver around it.	& [RL1.5.30]	"Instrument accuracy shall be reasonably checked prior to takeoff (e.g through regular inspection/calibration intervals). 
The lead aircraft shall monitor sensors for errors.
The lead aircraft shall perform a reasonableness check on the orientation angles before transmitting.
If the lead discovers a sensor error or unreasonable value for orientation angles, it shall be communicated as an invalid orientation rate in the position report.
The wingman aircraft shall perform a reasonableness check on the lead orientation angles. 
If the error is outside acceptable operational bounds, the wingman shall follow an appropriate contingency plan."	\\\hline

Lead does not provide its orientation angles to Wingman when the wingman is in close proximity to the lead.	&	The orientation angles might not be provided because of sensor error, calculation error, or recording error. If the wingman doesn't have the correct orientation angles for the lead, it won't be able to accurately keep track of the lead's position and navigation and won't be able to safely maneuver around it.	& [RL1.5.31]	"Instrument accuracy shall be reasonably checked prior to takeoff (e.g through regular inspection/calibration intervals). 
The lead aircraft shall monitor sensors for errors.
If the wingman uncertainty for the lead aircraft's orientation angles is outside acceptable operational bounds or cannot be estimated given a history of other data, the wingman shall follow an appropriate contingency plan."	\\\hline

\end{tabular}
\end{center}
\end{table}

\begin{table}[hbt!]
\begin{center}
\caption{\label{t:STPA_:L1510} Level 1 Signal 5: Lead to Wingman (10)}
\begin{tabular}[t]{|L{4.5cm} | L{5cm} | L{5.5cm} |}
\hline
\textbf{Unsafe Control Action}	&	\textbf{Rationale / Scenario}	&	\textbf{Requirement}	\\\hline

Lead provides the invalid value and details to Wingman when they are incorrect.	&	Error causes the invalid value to be incorrect. If the value is set to True when the position report is accurate, the wingman won't be able to trust the report and the flight may have to be cut short. If the value is set to False when the report is incorrect, the wingman will operate on incorrect information and won't have an accurate position for the lead	& [RL1.5.32]	"Instrument accuracy shall be reasonably checked prior to takeoff (e.g through regular inspection/calibration intervals). 
The lead aircraft shall monitor sensors for errors.
The lead aircraft shall perform a reasonableness check on the invalid value and details before transmitting.
If the lead discovers a sensor error or unreasonable value for invalid value and details, it shall be communicated as an invalid orientation rate in the position report.
The wingman aircraft shall perform a reasonableness check on the lead invalid value and details. 
If the error is outside acceptable operational bounds, the wingman shall follow an appropriate contingency plan."	\\\hline

Lead does not provide the invalid value and details to Wingman when the wingman is in close proximity to the lead.	&	Error causes the invalid value to be left off of the position report. This value is used to check the validity of the report, so the wingman will be unable to trust any of the information in the report and the flight may have to be cut short	& [RL1.5.33]	"Instrument accuracy shall be reasonably checked prior to takeoff (e.g through regular inspection/calibration intervals). 
The lead aircraft shall monitor sensors for errors.
If the wingman uncertainty for the lead aircraft's invalid value and details is outside acceptable operational bounds or cannot be estimated given a history of other data, the wingman shall follow an appropriate contingency plan."	\\\hline

Lead provides its wind velocities to Wingman when they are incorrect.	&	The wind velocities might be incorrect because of sensor error, calculation error, or recording error. If the wingman doesn't have the correct wind velocities for the lead, it won't be able to accurately keep track of the lead's position and navigation and won't be able to safely maneuver around it.	& [RL1.5.34]	"Instrument accuracy shall be reasonably checked prior to takeoff (e.g through regular inspection/calibration intervals). 
The lead aircraft shall monitor sensors for errors.
The lead aircraft shall perform a reasonableness check on the wind velocities before transmitting.
If the lead discovers a sensor error or unreasonable value for wind velocities, it shall be communicated as an invalid orientation rate in the position report.
The wingman aircraft shall perform a reasonableness check on the lead wind velocities. 
If the error is outside acceptable operational bounds, the wingman shall follow an appropriate contingency plan."	\\\hline

\end{tabular}
\end{center}
\end{table}

\begin{table}[hbt!]
\begin{center}
\caption{\label{t:STPA_:L1511} Level 1 Signal 5: Lead to Wingman (11)}
\begin{tabular}[t]{|L{4.5cm} | L{5cm} | L{5.5cm} |}
\hline
\textbf{Unsafe Control Action}	&	\textbf{Rationale / Scenario}	&	\textbf{Requirement}	\\\hline

Lead does not provide its wind velocities to Wingman when the wingman is in close proximity to the lead.	&	The wind velocities might not be provided because of sensor error, calculation error, or recording error. If the wingman doesn't have the correct wind velocities for the lead, it won't be able to accurately keep track of the lead's position and navigation and won't be able to safely maneuver around it.	& [RL1.5.35]	"Instrument accuracy shall be reasonably checked prior to takeoff (e.g through regular inspection/calibration intervals). 
The lead aircraft shall monitor sensors for errors.
If the wingman uncertainty for the lead aircraft's wind velocities is outside acceptable operational bounds or cannot be estimated given a history of other data, the wingman shall follow an appropriate contingency plan."	\\\hline

Lead provides voice coordination and safety concerns to Wingman when they are unnecessary.	&	This could be due to the lead having incorrect information or not knowing what information is and isn't necessary to share. This could be unsafe because it could be distracting to both pilots and might prevent necessary information from getting through. [CCS9, CCS12]	& [RL1.5.36]	"All flight personnel shall be briefed on communication protocol before takeoff. 
Communication systems shall be tested before takeoff to ensure they work."	\\\hline

Lead does not provide voice coordination and safety concerns to Wingman when they are necessary.	&	This could be due to the lead having incorrect information or not knowing what information is and isn't necessary to share. It also could be due to communications dropout. This is unsafe because the lead and the wingman are not coordinating	&[RL1.5.37]	"All flight personnel shall be briefed on communication protocol before takeoff. 
Communication systems shall be tested before takeoff to ensure they work and measures shall be taken to fix dropouts if possible. 
If the dropout can't be fix, a contingency plan shall be followed"	\\\hline

Lead does not provide voice coordination and safety concerns to Wingman when there is a legitimate safety concern.	&	This could be due to the lead having incorrect information, being unaware of unsafe situations, or communications dropout. It's unsafe because the wingman can't be made aware of the safety concern and coordination	& [RL1.5.38]	"All flight personnel shall be briefed on communication protocol before takeoff. 
All flight personnel shall also be trained to recognize unsafe situations and be in contact with someone else who can if they are unable to recognize unsafe situations. 
Communication systems shall be tested before takeoff to ensure they work and measures shall be taken to fix dropouts if possible. 
If the dropout can't be fix, a contingency plan shall be followed"	\\\hline

Lead provides voice coordination and safety concerns to Wingman too long when some of the information is unnecessary.	&	This could be due to the lead having incorrect information or not knowing what information is and isn't necessary to share. This could be unsafe because it could be distracting to both pilots and might prevent necessary information from getting through. [CCS9, CCS12]	& [RL1.5.39]	"All flight personnel shall be briefed on communication protocol before takeoff. 
Communication systems shall be tested before takeoff to ensure they work."	\\\hline

\end{tabular}
\end{center}
\end{table}

\begin{table}[hbt!]
\begin{center}
\caption{\label{t:STPA_:L1512} Level 1 Signal 5: Lead to Wingman (12)}
\begin{tabular}[t]{|L{4.5cm} | L{5cm} | L{5.5cm} |}
\hline
\textbf{Unsafe Control Action}	&	\textbf{Rationale / Scenario}	&	\textbf{Requirement}	\\\hline

Lead provides voice coordination and safety concerns to Wingman too short when not enough detail is provided.	&	This could be due to the lead having incorrect information or not knowing what information is and isn't necessary to share. This could be unsafe because the wingman wouldn't have enough information to understand and act on the concerns and coordination.	& [RL1.5.40]	"All flight personnel shall be briefed on communication protocol before takeoff. 
If the wingman pilot needs more detail, they shall request it from the lead if possible. 
Communication systems shall be tested before takeoff to ensure they work."	\\\hline

Lead provides voice coordination and safety concerns to Wingman too early when the information is not yet relevant.	&	This could be due to the lead having incorrect information or trying to coordinate prematurely. This could be unsafe because it could be distracting to both pilots and might prevent necessary information from getting through. [CCS9, CCS12]	& [RL1.5.41]	"All flight personnel shall be briefed on communication protocol before takeoff. 
If the lead is sharing information too early, they shall stop and notify the other parties, and then repeat the information at the appropriate time. 
If the wingman tries to act on premature information, an effort shall be made to correct the mistake."	\\\hline

Lead provides voice coordination and safety concerns to Wingman too late when the event has already passed.	&	This could be due to the lead being unaware of the event or a delay in communications.	& [RL1.5.42]	"All flight personnel shall be briefed on communication protocol before takeoff. 
If the lead fails to share information on time, the wingman shall make a best effort to maintain safe flight if necessary and a contingency plan shall be followed"	\\\hline

\end{tabular}
\end{center}
\end{table}

\end{document}